\documentclass[3p, 11pt]{elsarticle}
\usepackage{graphicx,amscd,amsmath,amssymb,verbatim}
\usepackage{amsfonts,epsfig}
\usepackage{mathptmx}
\usepackage{setspace}
\usepackage{epsfig}
\usepackage{url}
\usepackage{multirow}
\usepackage{subfigure}
\usepackage{color}

\begin{document}

\journal{Elsevier}

\begin{frontmatter}

\title{Randomness and Arbitrary Coordination in the Reactive Ultimatum Game}

\author{Roberto da Silva $^{1}$,  Pablo Valverde$^{1}$, Luis C. Lamb$^{2}$}

\address{1-Institute of Physics, Federal University of Rio Grande do Sul,
Av. Bento Gon\c{c}alves, 9500, Porto Alegre, 91501-970, RS, Brazil
{\normalsize{E-mail:rdasilva@if.ufrgs.br, pblvalverde@gmail.com}}}

\address{2-Institute of Informatics, Federal University of Rio Grande do Sul,
Av. Bento Gon\c{c}alves, 9500, Porto Alegre, 91501-970, RS, Brazil
{\normalsize{E-mail:lamb@if.ufrgs.br}}}


\begin{abstract}

The ultimatum game explains and is a useful model in the analysis of several effects of 
bargaining in population dynamics. Darwin's theory of evolution - as introduced in game theory 
by Maynard Smith -  is not the only important evolutionary  aspect in a evolutionary dynamics, 
since complex interdependencies, competition, and growth should be modeled by, for example, 
reactive aspects. In biological models, computationally or analytically considered, several authors 
have been able to show the emergence of cooperation with stochastic or deterministic dynamics based 
on the mechanism of copying the best strategies. On the other hand, in the ultimatum game the
reciprocity and the fifty-fifty partition seems to be a deviation from rational behavior of the players under the light 
of the Nash equilibrium concept.  Such equilibrium  emerges from the punishment of the responder who 
generally tends to refuse unfair proposals. In the iterated version of the game, the proposers are able to improve 
their proposals by adding an amount thus making fairer proposals. Such evolutionary aspects are not properly 
Darwinian-motivated, but they are endowed with a fundamental aspect: they reflect their actions according to 
value of the offers. Recently,  a reactive version of the ultimatum game where the acceptance occurs with fixed 
probability was proposed. In this paper, we aim at exploring this reactive version of the ultimatum game where 
the acceptance by the players depends on the offer. In order to do so, we analyze two situations: (i) mean field  and (ii) 
by considering the players inserted within the networks with arbitrary coordinations. In the proposed model we not only 
explore situations of occurrence of the fifty-fifty steady-state, in both homogeneous and heterogeneous populations, 
but also explore the fluctuations and  payoff distribution characterized by the Gini coefficient of the  population. 
We then show that the reactive aspect, here studied, thus far not analyzed in the evolutionary game theory literature 
can unveil an essential feature for the convergence to fifty-fifty split. Our approach concerns four different policies 
to be adopted by the players. In such policies the evolutionary aspects do not work through a Darwinian copying mechanism, 
but by following a policy that governs the increase or decrease of their offers according to the response of the 
result - i.e. acceptance or refusal. Moreover, we present results where the acceptance occurs with fixed probability. Our 
contribution is twofold: we present both analytical results and MC simulations which in turn are useful to design new controlled 
experiments in the ultimatum game in stochastic and deterministic scenarios.

\end{abstract}

\end{frontmatter}


\setlength{\baselineskip}{0.7cm}

\section{Introduction}

\label{section:introduction}

Game theory analyzes several important aspects of the Economical and
Biological sciences such as bargaining, cooperation and other social
features. The theory plays an important role in explaining the interaction
between individuals in homogeneous and heterogeneous populations, with or
without spacial structure, in which agents negotiate/combat/collaborate via
certain protocols. The full understanding of cooperation between individuals
as an emergent collective behavior remains an open challenge~\cite%
{Neumann,Smith,SzaboFath}. In this context, bargaining is an important
feature has called attention of many authors: two players must divide an
amount (resources, money, food, or other interesting quantity) and the
disagreement (or no agreement) between them in a given deal could mean that
both lose something. This dilemma motivates a simple game that mimics the
bargaining between two players - the Ultimatum Game.

In this game, firstly proposed by G\"{u}th \textit{et al.}~\cite{Guth1982},
one of the players proposes a division (the proposer) and the second player
(the responder) can either accept or reject it. If the responder (the second
player) accepts it, the values are distributed according to the division
established by the proposer. Otherwise, no earning is distributed to both
players.

Real situations in western societies suggest that unfair proposals are
refused for either fairer or even more selfish amounts. However some
isolated societies as Machiguenga localized at Peruvian Amazon seem to show
a behavior opposed to such fact, which suggests a more altruistic behavior 
\cite{Henrich2000}. On the other hand, scientists have studied and simulated
artificial societies where players confront each other according the
ultimatum game protocol. In order to consider a simple evolutionary
probabilistic model where unsatisfactory proposals are refused, in this
paper we propose to study a model where accepting depends on proposal. 
\footnote{%
This game scenario is common and expected in real situations, at least in
western societies, illustrated even when children negotiate chocolate coins
(see e.g. this video https://www.youtube.com/watch?v=YXfEv-xEWtE).}

Although it is rationally better for the responder to accept any offer,
offers below one third of the available amount to be shared are often
rejected \cite{juespacial}. The responder punishes the proposer up to the
balance between proposal and acceptance in the iterated game. In general,
values around a half of the total amount are accepted~\cite%
{juespacial,estudoneural}. Other interesting experimental results suggest
that high-testosterone men reject low offers in the ultimatum game \cite%
{Burnaham}. Nowak et. al \cite{Nowak2000} showed that the evolution of
fairness, similarly to the evolution of cooperation, is linked to reputation
by considering a simple memory mechanism: fairness will evolve if the
proposer can obtain some information on what deals the responder has
accepted in the past .

Our contribution goes precisely along this line of research. In this
manuscript, we extend the memory-1 model proposed by one of the authors in 
\cite{Enock2014} that considers the acceptance with fixed probability, by
putting this probability variable and assigning the offer $O_{t}$, at time $%
t $, that is a number belonging to $[0,1]$ and performing the game in graphs
with arbitrary homogeneous and heterogeneous coordination.

In this reactive and iterated version of the ultimatum game, the players are
able to correct their offers by adding/subtracting an amount to the offers
in order to make fairer proposals. Such mechanisms, which we assume are an
essential ingredient for the convergence to fifty-fifty partitions seems to
be discarded in typical evolutionary game theory based on probabilistic
Darwinian copies. By performing a detailed study, we investigate the game
both analytically and via Monte Carlo (MC) simulations under four different
policies about the increase or decrease of the offer under different levels
of greed. Moreover, we present results about temporal correlations in the
model with fixed probability for a suitable comparison with the model where
the offer is time-dependent.

The remainder of paper is organized as follows. Next, we define the reactive
model and its mean-field approximation. Then, we show how the model can be
run in networks with arbitrary coordination. In Section \ref%
{Section:Mean-field} we present the first part of our results corresponding
to the mean-field approximation. In Section \ref{Section:Results-II} we
present the results for the game with arbitrary coordination via equation
integrations. Particularly for $k=4$ we explore the randomness effects by
considering MC simulations in small world networks. A general and analytical
formula is obtained for the stationary average offer and a complete study of
the fluctuations and distribution of the payoffs are performed considering
homogeneous and heterogeneous populations. Then we present a comparative
analysis between mean-field and the model on networks. Finally, we conclude
and comment on the relevance of the reactive ultimatum game, in particular
on the experimental evidence of the effect of fairer offers in different
international societies.

\section{Modeling and Mean-field Approximation: Analyzing the correlations}

\label{Section:Mean-field}

In the reactive ultimatum game, when a player (proposer) performs an offer $%
O_{t}\in \lbrack 0,1]$ at time $t$, it can b accepted or rejected by the
other player (i.e. the responder). Let us think that such acceptance occurs
with probability $p_{t}$. Let us consider two simple situations:

\begin{enumerate}
\item $p_{t}=p$ fixed, and does not change along time;

\item $p_{t}=O_{t}$, i.e., the acceptance occurs with higher probability as
the offer is more generous.
\end{enumerate}

When the offer is rejected it will take the proposer to change its
expectations increasing its proposed offer $\epsilon$. On the other hand
when it is accepted the proposer decreases its proposal by a quantity $%
\epsilon$. Here $\epsilon $ is a rate of offer change. We can consider the
mean-field regime as the average under all different time series of
parameters of two players interacting according to a dynamics. We also can
imagine it as parameters averaged by the different players in a large
population, where the players interact at each time $t$ (denoted by authors
in refs. \cite{SilvaJTB}, and \cite{SIlvaBJP} as one `turn') by pairs
composing a perfect matching with $N$ players (for the sake of simplicity $N$
is an even number) randomly composed. In this pairing, no player is left out
of the game, with each individual playing once by turn, by construction.

Both ways provide similar ways to compute averaged parameters evolving along
time, since in this reactive formulation of the ultimatum game, the
interaction depends only on the proposal (offer). The first case ($p$ fixed)
were partially explored in \cite{Enock2014} but some important points
involving the existing correlations, have not been studied yet.

First, we would like to revisit the problem $p_{t}=p$ to describe the
possible correlations which were not studied in \cite{Enock2014}. In this
case the clustering effects are not important, and in the next subsection we
revisit some results for $p_{t}=p$ to deduce some semi-analytical formulas
for the sum of temporal correlations of the payoff. Next, in the following
subsection, we define the model for $p_{t}=O_{t}$, and we deduce some
relevant results by mean-field approximation. Our results show that
independently from $O_{0}$, $\lim_{t\rightarrow \infty }O_{t}=1/2$.

\subsection{Reactive Ultimatum Game With $p_{t}=p$: Mean-field approximations%
}

Let us consider the case where the responder always accepts the offer with a
fixed probability $p\in \lbrack 0,1]$ \cite{Enock2014}, and the offer
rejection occurs with probability $1-p$. This assumption allows us to obtain
analytical results in the one-step memory iterated game. Given $\varepsilon $
and $p$, in the $i-$th round, the average offer is: 
\begin{equation}
\left\langle O_{i}\right\rangle =O_{0}+i\varepsilon (1-2p),
\label{Eq:average_offer}
\end{equation}%
where $i=0,1,2\ldots t$, since in each round the average offer is modified
by $\left\langle \left( \bigtriangleup O\right) _{i}\right\rangle
=(1-p)\varepsilon -p\varepsilon =$ $\varepsilon (1-2p)$. In the $i$-th
round, the responder average payoff is $\left\langle g_{i}\right\rangle
=p\left\langle O_{i}\right\rangle =pO_{0}+ip\varepsilon (1-2p)$. Thus, after 
$t$ iterations, the average of the cumulative payoff is 
\begin{equation}
\left\langle W_{r}\right\rangle (t)=\sum\limits_{i=0}^{t}\left\langle
g_{i}\right\rangle =pO_{0}(t+1)+\frac{t(t+1)}{2}p(1-2p)\varepsilon
\label{Eq:Acumulado_p_fixo}
\end{equation}
and there is a probability $p$, for a given $n$, that maximizes the
cumulative responder gain $\left\langle W_{r}\right\rangle (t)$ is given by $%
p^{\ast }=\frac{1}{4}\left[ \frac{2y_{0}}{t\bigtriangleup y}+1\right] $.
Similarly we have that for proposer the average cumulative payoff is given
by $\left\langle W_{p}\right\rangle (t)=p(1-O_{0})(t+1)-\frac{t(t+1)}{2}%
p(1-2p)\varepsilon $.

In order to calculate the variance of the cumulative gain, the task is not
so simple. The result was obtained in \cite{Enock2014} but only this
computed result was shown. Basically, this is not only an analytical task.
We suppose that variance is four-degree polynomial $p$ with at least two
roots: $p=0$ and $p=1$. So the variance is considered as a polynomial $%
var(W_{r})=ap(p-1)(p-p_{1})(p-p_{2})$ where $a$, $p_{1}$ and $p_{2}$ are
constants to be determined. By observing the variance for an arbitrary
number of rounds (numerically) for three different $p$ values $p=1/2$, $%
p=1/4 $ and $p=3/4$ we solve a linear system to find $a$, $p_{1}$ and $p_{2}$
and we can check the semi-empirical analytical formula obtained in \cite%
{Enock2014}:

\begin{equation}
\begin{array}{lll}
var(W_{r})(t) & = & (t+1)p(1-p)O_{0}^{2}+4t(t+1)p(p-1)\left( p-\frac{1}{4}%
\right) O_{0}\varepsilon +[2t(t+1)(2t-1)p^{3}(1-p) \\ 
&  &  \\ 
&  & -2t(t^{2}-1)p^{2}(1-p)+\frac{t(t+1)(2t+1)p(1-p)}{6}]\varepsilon ^{2}%
\end{array}
\label{Eq:varproposer}
\end{equation}
and similarly, we can obtain the variance of the cumulative gain of the
proposer: $var(W_{p})=(t+1)p(1-p){(1-O_{0})}^{2}-4t(t+1)p(p-1)\left( p-\frac{%
1}{4}\right) (1-O_{0})\varepsilon
+[2t(t+1)(2t-1)p^{3}(1-p)-2t(t^{2}-1)p^{2}(1-p)+\frac{t(t+1)(2t+1)p(1-p)}{6}%
]\varepsilon ^{2}$.

Implicitly, our difficulty in analytically obtaining a formula to the
variance of the gain is related to the the fact that there is no control of
correlation in the problem. Here aim at providing a more detailed
exploration in order to understand the correlations involved in such a
problem.

Since for example $var(W_{r})(t)=\sum_{t^{\prime }=0}^{t}\left\langle
g_{t^{\prime }}^{2}\right\rangle -\left\langle g_{t^{\prime }}\right\rangle
^{2}+\sum_{t^{\prime }=0}^{t}\sum_{t^{\prime \prime }=0}^{t}\left(
\left\langle g_{t^{\prime }}g_{t^{\prime \prime }}\right\rangle
-\left\langle g_{t^{\prime }}\right\rangle \left\langle g_{t^{\prime \prime
}}\right\rangle \ \right) =$ $\sum_{t^{\prime }=1}^{t}var(g_{t^{\prime
}})+\sum_{t^{\prime }=1}^{t}\sum_{t^{\prime \prime }=1}^{t}corr(g_{t^{\prime
}},g_{t^{\prime \prime }})$. Let us think about the first part of sum: we
can write that $\left\langle g_{t^{\prime }}^{2}\right\rangle =p\left\langle
O_{t^{\prime }}^{2}\right\rangle $. But, how can one compute $\left\langle
O_{t^{\prime }}^{2}\right\rangle $? Since $\left\langle O_{t^{\prime
}}{}^{2}\right\rangle =(1-p)\left\langle \left( O_{t^{\prime
}-1}+\varepsilon \right) ^{2}\right\rangle +p\left\langle \left(
O_{t^{\prime }-1}-\varepsilon \right) ^{2}\right\rangle $, we have that $%
\left\langle O_{t^{\prime }}{}^{2}\right\rangle =\varepsilon
^{2}+\left\langle O_{t^{\prime }-1}^{2}\right\rangle +2\varepsilon
(1-2p)\left\langle O_{t^{\prime }-1}\right\rangle $. We can easily conclude,
by iterating such equation, that: $\left\langle O_{t}{}^{2}\right\rangle
=O_{0}^{2}+2(1-2p)O_{0}t\epsilon +(t(t-1)(1-2p)^{2}+t)\epsilon ^{2}$. So $%
\left\langle g_{t}^{2}\right\rangle =p[O_{0}^{2}+2(1-2p)O_{0}t\epsilon
+(t(t-1)(1-2p)^{2}+t)\epsilon ^{2})]$ and $\left\langle g_{t}\right\rangle
^{2}=p^{2}\left[ O_{0}^{2}+2(1-2p)O_{0}t\epsilon +(1-2p)^{2}t^{2}\epsilon
^{2}\right] $. Expanding the terms we have that 
\begin{eqnarray}
\left\langle g_{t}^{2}\right\rangle -\left\langle g_{t}\right\rangle ^{2}
&=&p(1-p)O_{0}^{2}+2p(1-p)(1-2p)O_{0}t\epsilon  \label{local_gain_variance}
\\
&&+[p(t(t-1)(1-2p)^{2}+t)-p^{2}(1-2p)^{2}t^{2}]\epsilon ^{2}  \notag
\end{eqnarray}

By performing the sum we obtain:

\begin{eqnarray}
\overline{var(W_{r})}(t) &=&\sum_{t^{\prime }=0}^{t}\left\langle
g_{t^{\prime }}^{2}\right\rangle -\left\langle g_{t^{\prime }}\right\rangle
^{2}=\left( 2p^{2}(1-p)t(t+1)+\frac{1}{6}p(1-p)(1-2p)^{2}t\left( 2t+1\right)
\left( t+1\right) \right) \epsilon ^{2}  \label{Eq:without_correlations} \\
&&+p(1-p)(1-2p)O_{0}t(t+1)\epsilon +p(1-p)O_{0}^{2}(t+1)  \notag
\end{eqnarray}

This formula, can be used to estimate the magnitude of correlations since
from Eq. \ref{Eq:varproposer} we have an exact form (empirically obtained)
for the variance. So by measuring this magnitude we can define the
following: 
\begin{equation}
\Phi (t)=\sum_{t^{\prime }=0}^{t}\sum_{t^{\prime \prime
}=0}^{t}corr(g_{t},g_{t^{\prime }})  \label{Eq:fi}
\end{equation}%
By some algebra derivations we obtain:

\begin{eqnarray}
\Phi (t) &=&2t(t+1)p^{2}(p-1)O_{0}\varepsilon +  \label{Eq:fi_p_fixo} \\
&&\left( \allowbreak -\frac{8}{3}p^{4}t^{3}+\frac{8}{3}p^{4}t+\frac{10}{3}%
p^{3}t^{3}-\frac{10}{3}p^{3}t-\frac{2}{3}p^{2}t^{3}+\allowbreak \frac{2}{3}%
p^{2}t\right) \varepsilon ^{2}  \notag
\end{eqnarray}

So we can study this function in detail. Since the offer $O_{i}$ does not
touch the limits ($0$ or $1$) there is a lower bound for the number of
iterations necessary for the system to reach such limits: $n_{c}=\min \left(
\left\lfloor y_{0}/\varepsilon \right\rfloor ,\left\lfloor
(1-y_{0})/\varepsilon \right\rfloor \right) $.

\subsection{Reactive Ultimatum Game With $p_{t}=O_{t}$: Mean-field
approximations}

In more realistic situations, the acceptance depends on the offer. So, a
natural choice is setting the accepting probability as exactly the value of
the offer. In this case, considering a simple "mean-field" approximation
where we change $O_{t}$ by $\left\langle O_{t}\right\rangle $, a recurrence
relation for the offer can be written as:

\begin{equation*}
\begin{array}{lll}
\left\langle O_{t+1}\right\rangle & = & \left\langle O_{t}\right\rangle
-\left\langle O_{t}\right\rangle \epsilon +(1-\left\langle
O_{t}\right\rangle )\epsilon \\ 
&  &  \\ 
& = & (1-2\epsilon )\left\langle O_{t}\right\rangle +\epsilon%
\end{array}%
\end{equation*}

By iterating this equation we obtain: 
\begin{equation}
\begin{array}{lll}
\left\langle O_{t}\right\rangle & = & (1-2\epsilon )^{t}O_{0}+\epsilon
\sum_{k=0}^{t-1}(1-2\epsilon )^{k} \\ 
&  &  \\ 
& = & \left( O_{0}-1/2\right) (1-2\epsilon )^{t}+1/2%
\end{array}
\label{Eq:offer_depends_on_time}
\end{equation}%
and $\lim_{t\rightarrow \infty }\left\langle O_{t}\right\rangle =1/2$. Since 
$(1-2\epsilon )^{t}=1-2\epsilon t+O(\epsilon ^{2})$ for intermediate $t$%
-values, since $\epsilon $ is a small number we have the asymptotical
behavior: 
\begin{equation}
\left\langle O_{t}\right\rangle \sim \left\{ 
\begin{array}{lll}
\left( O_{0}-1/2\right) (1-2\epsilon t)+1/2 &  & t\rightarrow 0 \\ 
&  &  \\ 
1/2 &  & t\rightarrow \infty%
\end{array}%
\right.  \label{Eq:oferta_assintotico}
\end{equation}

Therefore an approximation for the average gain of the responder at time $t$
is $\left\langle g_{t}\right\rangle \approx \left\langle O_{t^{\prime
}}\right\rangle ^{2}=\left( O_{0}-1/2\right) ^{2}(1-2\epsilon
)^{2t}+1/4+\left( O_{0}-1/2\right) (1-2\epsilon )^{t}$. That asymptotically
gives 
\begin{equation*}
\left\langle g_{t}\right\rangle \sim \left\{ 
\begin{array}{lll}
O_{0}^{2}+(2-4O_{0})\epsilon O_{0}t &  & t\rightarrow 0 \\ 
&  &  \\ 
\left( O_{0}-1/2\right) ^{2}+\left( O_{0}-1/2\right) +1/4 &  & t\rightarrow
\infty%
\end{array}%
\right.
\end{equation*}

In our approximation, in this Pavlovian version the offer must converge to a
fair proposal. This result although simple, deserves a lot of discussion in
the literature and distortions of this behavior must be better understood
since it has an important role in the Pavlovian version of the ultimatum
game.

So a formula for the average of the cumulative gain at time $t$ in mean
field approximation can be written, since the acceptance probability is the
owner's offer value:

\begin{equation}
\begin{array}{lll}
\left\langle W_{r}(t)\right\rangle & = & \sum_{t^{\prime
}=0}^{t}\left\langle O_{t^{\prime }}\right\rangle ^{2} \\ 
&  &  \\ 
& = & \left( O_{0}-\frac{1}{2}\right) ^{2}\frac{\left( 2\epsilon -1\right)
^{2t+2}-1}{4\epsilon \left( \epsilon -1\right) }+\left( O_{0}-\frac{1}{2}%
\right) \frac{1-\left( 1-2\epsilon \right) ^{t+1}}{2\epsilon }+\frac{t+1}{4}%
\end{array}
\label{Eq. Ganho_dep_oferta}
\end{equation}

Again, we have two regimes: for $t\rightarrow 0$, asymptotically we have $%
\left\langle W_{r}(t)\right\rangle \sim \left( O_{0}-\frac{1}{2}\right) ^{2}%
\frac{(t+1)}{1-\epsilon }+(t+1)+\frac{t+1}{4}=\left( \frac{5}{4}+\frac{1}{%
1-\epsilon }\left( O_{0}-\frac{1}{2}\right) ^{2}\right) (1+t)$. For $%
t\rightarrow \infty $, $\left\langle W_{r}(t)\right\rangle \sim $ $\left(
O_{0}-\frac{1}{2}\right) ^{2}\frac{1}{4\epsilon \left( 1-\epsilon \right) }+%
\frac{1}{2\epsilon }\left( O_{0}-\frac{1}{2}\right) +\frac{t+1}{4}\ $which
determines a crossover between two different linear behaviors.

If we extract the correlations, the variance of the cumulative gain:

\begin{equation*}
\begin{array}{lll}
\overline{var(W_{r})}(t) & = & var(W_{r})(t)-\sum_{t^{\prime
}=0}^{t-1}\sum_{t^{\prime \prime }=0}^{t-1}\left( \left\langle O_{t^{\prime
}}O_{t^{\prime \prime }}\right\rangle -\left\langle O_{t^{\prime
}}\right\rangle \left\langle O_{t^{\prime \prime }}\right\rangle \ \right)
\\ 
&  &  \\ 
& = & \sum_{t^{\prime }=0}^{t-1}\left\langle g_{t^{\prime
}}^{2}\right\rangle -\left\langle g_{t^{\prime }}\right\rangle ^{2}\text{.}%
\end{array}%
\end{equation*}

\begin{equation*}
\left\langle O_{t}{}^{2}\right\rangle \approx (1-\left\langle
O_{t-1}\right\rangle )\left\langle \left( O_{t-1}+\varepsilon \right)
^{2}\right\rangle +\left\langle O_{t-1}\right\rangle \left\langle \left(
O_{t-1}-\varepsilon \right) ^{2}\right\rangle
\end{equation*}

After some algebraic calculations: 
\begin{equation*}
\left\langle O_{t}{}^{2}\right\rangle =\varepsilon ^{2}+\left\langle
O_{t-1}^{2}\right\rangle +2\varepsilon \left\langle O_{t-1}\right\rangle
-4\varepsilon \left\langle O_{t-1}\right\rangle ^{2}\ 
\end{equation*}%
what after the iteration and some algebra leads to:%
\begin{equation}
\begin{array}{lll}
\left\langle O_{t}{}^{2}\right\rangle & = & \left\langle
O_{t-1}{}^{2}\right\rangle +\varepsilon ^{2}+2\varepsilon \left[
(O_{0}-1/2)\left( 1-2\epsilon \right) ^{t-1}+1/2\right] -4\varepsilon \left[
\left( 1-2\epsilon \right) ^{t-1}+1/2\right] ^{2} \\ 
&  &  \\ 
& = & O_{0}{}^{2}+\varepsilon ^{2}t-(O_{0}-1/2)\left[ 1-\left( 1-2\epsilon
\right) ^{t}\right] -\frac{(O_{0}-1/2)^{2}}{\left( 1-\epsilon \right) }\left[
1-\left( 1-2\epsilon \right) ^{2t}\right]%
\end{array}
\label{Eq:O2}
\end{equation}

Following exactly what we considered previously, we can approximate%
\begin{equation*}
\begin{array}{lll}
\left\langle g_{t^{\prime }}^{2}\right\rangle & \approx & \left\langle
O_{t^{\prime }}\right\rangle \left\langle O_{t^{\prime }}^{2}\right\rangle 
\text{ } \\ 
&  &  \\ 
\left\langle g_{t^{\prime }}\right\rangle ^{2} & \approx & \left\langle
O_{t^{\prime }}\right\rangle ^{4}%
\end{array}%
\end{equation*}

It is important to see that%
\begin{equation}
var(g_{t})=\left\langle g_{t^{\prime }}^{2}\right\rangle -\left\langle
g_{t^{\prime }}\right\rangle ^{2}\sim \frac{1}{2}(O_{0}{}^{2}+\varepsilon
^{2}t-(O_{0}-1/2)-\frac{(O_{0}-1/2)^{2}}{\left( 1-\epsilon \right) })-\frac{1%
}{16}  \label{Eq.local_dependent_offer_var}
\end{equation}
for $t\rightarrow \infty $, which leads to a linear behavior in time for the
variance differently from case where accepting occurs with fixed
probability. In this case $var(g_{t})$ has a quadratic leader term in time.

We can evaluate numerically the expression 
\begin{equation}
\overline{var(W_{r})}\approx \sum_{t^{\prime }=0}^{t}\left\langle
O_{t^{\prime }}\right\rangle \left( \ \left\langle O_{t^{\prime
}}^{2}\right\rangle -\left\langle O_{t^{\prime }}\right\rangle ^{3}\right)
\label{Eq. Desc_dep_offer}
\end{equation}
and naturally to compute $\Phi (t)$ as performed for the case of the fixed $%
p $ (equation \ref{Fig:fi_with_fixed_p}) for the particular case where
acceptance depends on the offer, but for this case we have to compute $%
var(W_{r})$ numerically by a Monte Carlo simulation differently from the
case where the acceptance occurs with fixed value of $p$ (\ref%
{Eq:varproposer}) and computing $\overline{var(W_{r})}$ by using \ref{Eq:O2}.

\section{Extending the Model to Networks}

In this second part we analyze the model considering coordination and
randomness. In this case we consider that players are inserted into a
network (or graph) by considering the reactive ultimatum game with
acceptance probability equal to offer $O_{t}$.

To extend our results to networks, we consider four different policies that
governs the update dynamics of the player offers in the network, which works
as a greedy level. Here, the term \textit{conservative} must be understood
by the policy: \emph{if you are not sure about the acceptance of your offer
in the neighborhood, you will increase your offer; otherwise you will
decrease it.}

Our simulations consider a simple initial condition: first, an initial offer 
$O_{0}$ is assigned equally to all players. Such initial condition is
initially adopted for the sake of simplicity.

At $t-$th simulation step, each player $i=1,...,N$ in the network, where $N$
is the number of nodes, offers a value for its $k_{i}$ neighbors. Each
neighbor accepts or not the proposal with probability $p_{a}(t)=O_{t}^{(i)}$%
, where $O_{t}^{(i)}$ is the offer of $i$-th player at time $t$. Since we
compute the number of players that accept the proposal, $n_{a}(i)$, we have
the possible policies:

\begin{enumerate}
\item \textbf{Conservative: }\textit{Ensures that more than half of the
neighbors accept the proposal in order to reduce the offer}- If $%
n_{a}(i)>k_{i}/2$, so $O_{t+1}^{(i)}=O_{t}^{(i)}-\varepsilon $, otherwise $%
O_{t+1}^{(i)}=O_{t}^{(i)}+\varepsilon $;

\item \textbf{Greedy: }\textit{One acceptance is enough to reduce the offer}
- If $n_{a}(i)\geq 1$, so $O_{t+1}^{(i)}=O_{t}^{(i)}-\varepsilon $,
otherwise $O_{t+1}^{(i)}=O_{t}^{(i)}+\varepsilon $;

\item \textbf{Highly Conservative: }\textit{All neighbors must accept the
proposal to reduce the offer} - If $n_{a}(i)=k_{i}$, so $%
O_{t+1}^{(i)}=O_{t}^{(i)}-\varepsilon $, otherwise $%
O_{t+1}^{(i)}=O_{t}^{(i)}+\varepsilon $;

\item \textbf{Moderate}: If exactly half of the neighbors accept it, then
the proposal is reduced - $n_{a}(i)\geq k_{i}/2$, , so $%
O_{t+1}^{(i)}=O_{t}^{(i)}-\varepsilon $, otherwise $%
O_{t+1}^{(i)}=O_{t}^{(i)}+\varepsilon $;
\end{enumerate}

Let us consider a particular and interesting case, where the coordination of
all nodes is fixed and made equal to $k$ (regular graph). For example, in
the first case we have, 
\begin{equation*}
\begin{array}{lll}
\left\langle O_{t+1}\right\rangle & \approx & \left\langle
O_{t}\right\rangle -\varepsilon \left( \Pr (n_{a}>k/2|\left\langle
O_{t}\right\rangle )-\Pr (n_{a}\leq k/2|\left\langle O_{t}\right\rangle
)\right) \\ 
&  &  \\ 
& = & \left\langle O_{t}\right\rangle +\varepsilon \left( 2\Pr (n_{a}\leq
k/2|\left\langle O_{t}\right\rangle )-1\right)%
\end{array}%
\end{equation*}

But 
\begin{equation*}
\Pr (n_{a}\leq k/2|\left\langle O_{t}\right\rangle )=\sum_{m=0}^{k/2}\frac{%
k!\left\langle O_{t}\right\rangle ^{m}(1-\left\langle O_{t}\right\rangle
)^{k-m}}{m!(k-m)!}
\end{equation*}%
and so

\begin{equation}
\left\langle O_{t+1}\right\rangle \approx \left\langle O_{t}\right\rangle + 
\left[ 2\sum_{m=0}^{k/2}\frac{k!\left\langle O_{t}\right\rangle
^{m}(1-\left\langle O_{t}\right\rangle )^{k-m}}{m!(k-m)!}-1\right]
\varepsilon  \label{Eq:Exact_recurrence}
\end{equation}

We can iterate this recurrence relation and compare with results from Monte
Carlo simulations in networks with fixed coordination $k$. Monte Carlo
simulations can also be performed to analyze the deviations of this formula
when the average degree is $k$ in disordered networks. In section \ref%
{Section:Results-II} we analyze, for example, the deviations from formula %
\ref{Eq:Exact_recurrence} when we introduce effects of randomness $p$ in
small worlds built from rings and two-dimensional lattices.

In this same section we present studies about payoff distribution for $k=4$
and analyses of stationary offer $\left\langle O_{\infty }\right\rangle $
for arbitrary $k$ in heterogeneous population of players, i.e., we consider
different partition of players that play under four different policies.

\section{Results Part I: Mean-field Regime}

In the sequel, we present our main results in the mean-field regime.

\subsection{Mean-field for acceptance with fixed probability $p$}

In the previous section, we observe that in such case, the offer increases
or decreases linearly with time. The cumulative payoff (wealth) of the
responder ($\left\langle W_{r}\right\rangle (t)$) also is easily calculated
by Eq. \ref{Eq:Acumulado_p_fixo}. For $p=1/2$, we can verify that $%
\left\langle W_{r}\right\rangle $ grows linearly in time, independently from
rate $\varepsilon $. The quadratic term is relevant for $p\neq 1/2$. This
simple calculation suggests that the variance of cumulative gain should also
be calculated. Here a problem occurs: The authors in \cite{Enock2014} have
analyzed this particular case of the game. They calculated it by using a
semi-empirical method to obtain $var(W_{r})(t)$, fitting a polynomial in $p$
which has two obvious roots: $p=0$ and $p=1$, resulting in equation \ref%
{Eq:varproposer}. This is so because the authors avoid the correlations of
the problem, the only reason that prohibits an analytical derivation of this
formula by direct methods. But why is this important? Because we can use the
semi-empirical formula for the variance obtained by \cite{Enock2014} given
by Equation \ref{Eq:varproposer} in order to study the correlations of the
problem.

By performing such correlations, first of all, it is important to observe
the behavior of variances of the payoff at time $t$ of responder: $%
var(g_{t})=\left\langle g_{t}^{2}\right\rangle -\left\langle
g_{t}\right\rangle ^{2}$. We know how to calculate this value as can be seen
in Eq. \ref{local_gain_variance}. Here our first study is to observe the
influence of the changing rate of $\varepsilon $. \ 

\begin{figure}[th]
\begin{center}
\includegraphics[width=%
\columnwidth]{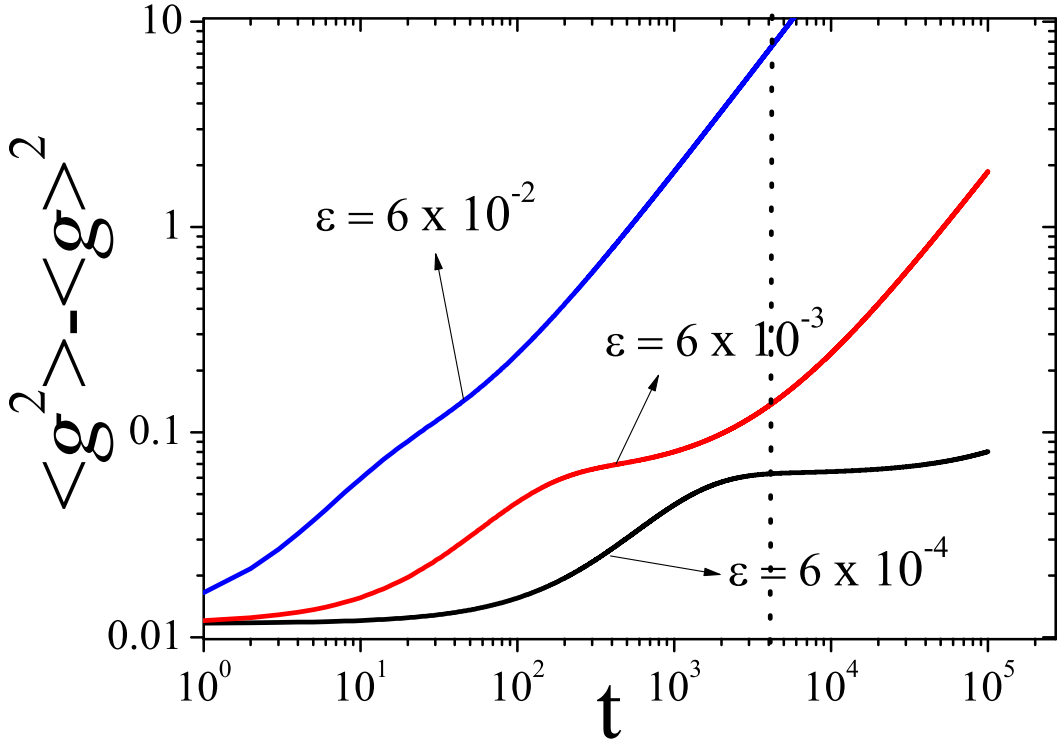}
\end{center}
\caption{Variance of the payoff of responder for $p=1/2$ and $O_{0}=1/4$,
for different $\protect\varepsilon $-values according to Eq. \protect\ref%
{local_gain_variance} We can observe in this log-log plot different stages
of growing as function of time. }
\label{Fig:epsilon_effects_p_fixed}
\end{figure}

We can observe that $var(g_{t})$ increases with time, as Fig. \ref%
{Fig:epsilon_effects_p_fixed}. A convexity change is more sensitive to
higher $\varepsilon $-values. The sum of these local variances corresponds
to a part of $var(W_{r})(t)$, denoted by $\overline{var(W_{r})}(t)$ which
was analytically obtained by Eq. \ref{Eq:without_correlations}. So we
compute $\Phi (t)=var(W_{r})(t)-\overline{var(W_{r})}(t)$ estimated by
equation \ref{Eq:fi_p_fixo} which corresponds to the sum of all correlations
of the payoffs until time $t$, i.e., a kind of \textquotedblleft
cumulative\textquotedblright\ correlation.

\begin{figure}[th]
\begin{center}
\includegraphics[width=%
\columnwidth]{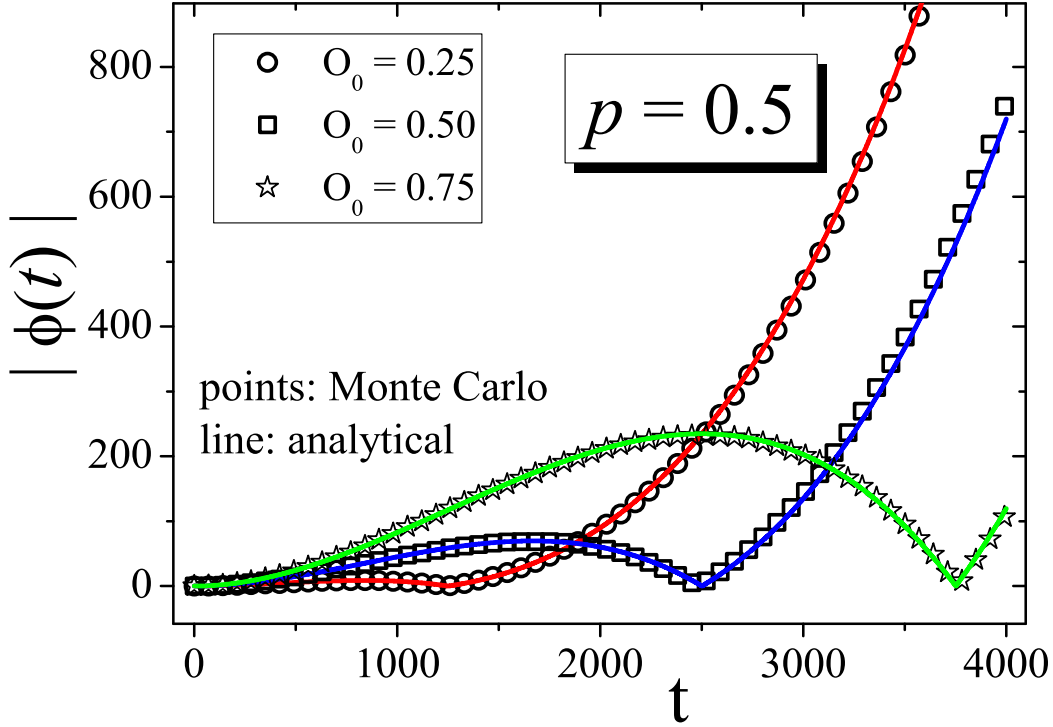}
\end{center}
\caption{Absolute value of $\Phi (t)$ given by Eq. \protect\ref{Eq:fi_p_fixo}
for different values of $O_{0}$, with $p=1/2$. We can observe that there is
a $t^{\ast }$ which corresponds to $\Phi =0$ for each $O_{0}$. Such point
corresponds to a signal change of $\Phi (t)$. }
\label{Fig:fi_with_fixed_p}
\end{figure}

In Fig. \ref{Fig:fi_with_fixed_p} we can check the behavior of $\left\vert
\Phi (t)\right\vert $. The points $t^{\ast }(O_{0})$ where $\Phi =0$, give $%
var(W_{r})(t^{\ast })=\overline{var(W_{r})}(t^{\ast })$, i.e, they work as a
\textquotedblleft decorrelation time\textquotedblright\ of the system that
depends on initial offer $O_{0}$. The points in Fig. \ref%
{Fig:fi_with_fixed_p} corresponds to MC simulations used to corroborate the
results from equation \ref{Eq:fi_p_fixo}. In this simulations we performed $%
10^{5}$ runs of the iterated game performing averages for each time. We can
see a perfect agreement between Eq. \ref{Eq:fi_p_fixo} and MC simulations.

Now let us show the results for the reactive ultimatum game version when the
acceptance depends on the offer and compare with the results of this
subsection.

\subsection{Mean-field for Acceptance Dependent on the Offer}

We observed that reactive $p$-fixed approach for acceptance of the offer
leads to offers that increase or decrease along time. This is a possible
behavior, but the experiments with human beings (see e.g. \cite{Henrich2000}%
) seem to avoid the undesirable situation leading to a fair steady state:
fifty-fifty sharing.

The reactive ultimatum game, based on acceptances that depend on the offers
produce a stable state $O_{\infty }=1/2$ independently of $O_{0}$. This
strong fixed point must be better understood. Here, the first question is to
check the cumulative gain and variance of the offer for $O_{0}=1/2$ in this
situation comparing these same values in the reactive $p$-fixed game.

\begin{figure}[th]
\begin{center}
\includegraphics[width=0.5\columnwidth]{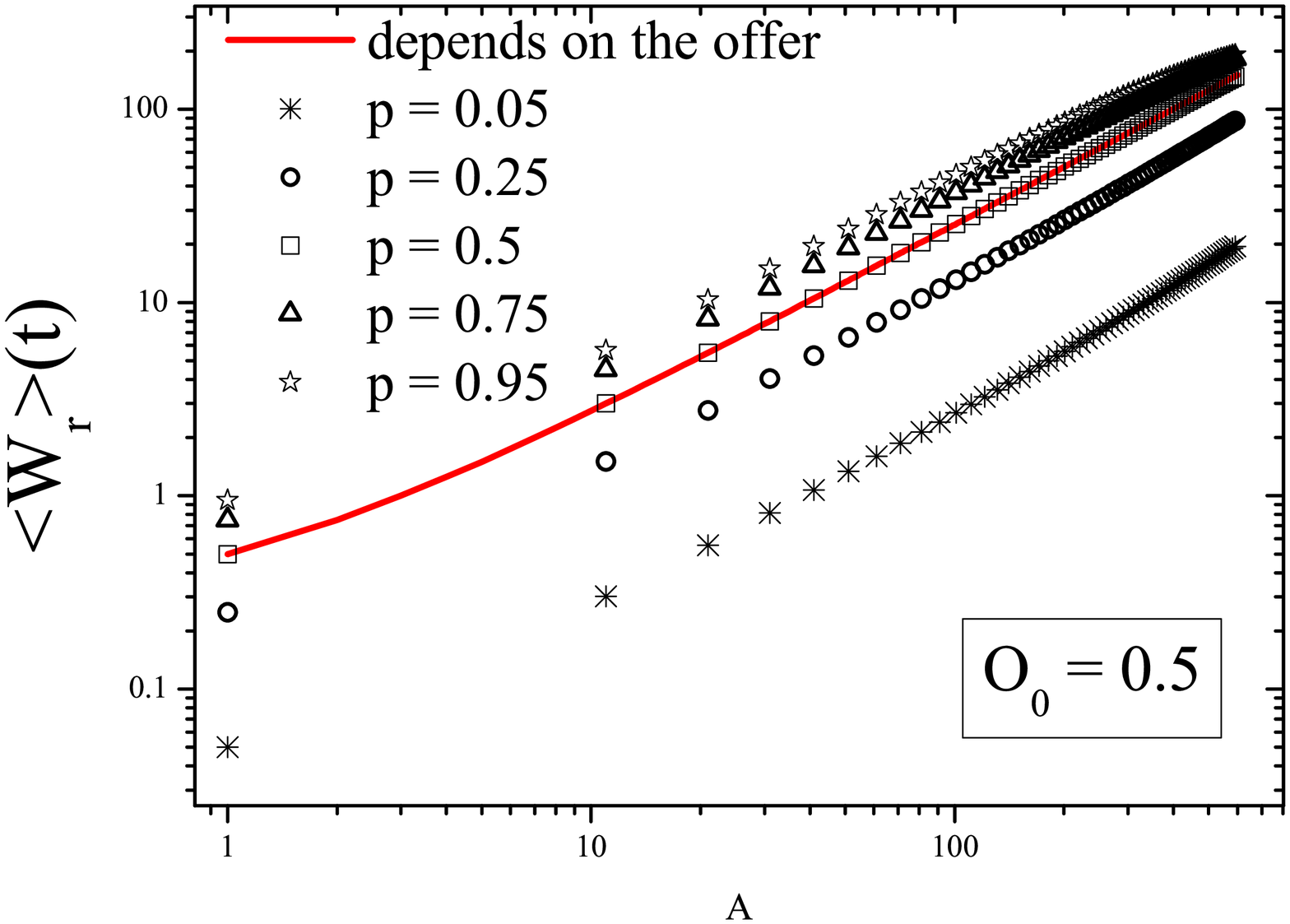}%
\includegraphics[width=0.5%
\columnwidth]{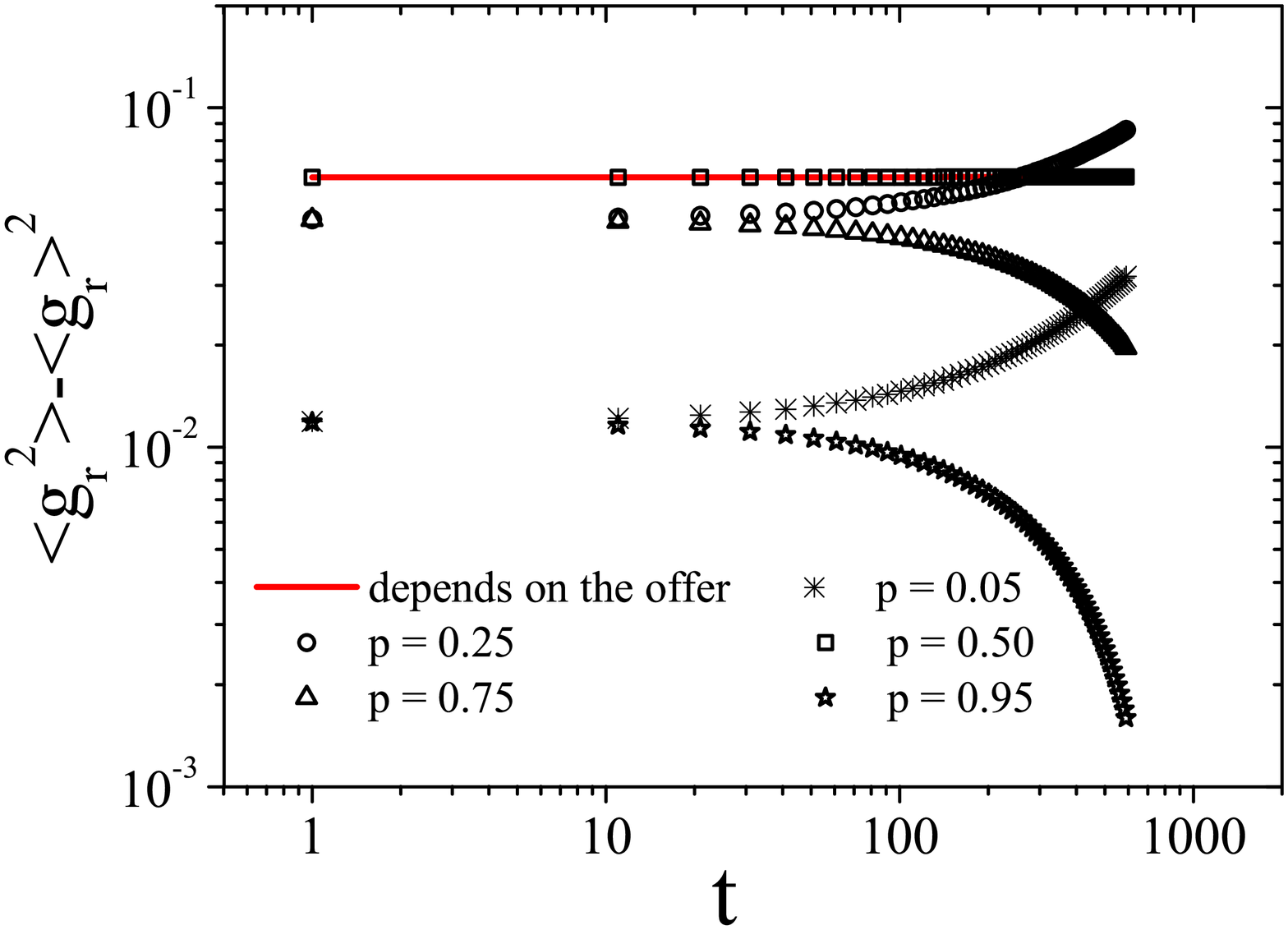}
\end{center}
\caption{\textbf{Left}: Cumulative payoff (wealth) for different $p-$values
with $O_{0}=1/2$. The red line corresponds to the case in which accepting
depends on the offer. It fits very well the case $p=1/2$. \textbf{Right}:
The same results corresponding to the variance of payoff. }
\label{Fig:ganho_variance}
\end{figure}

In Figure \ref{Fig:ganho_variance} (\textbf{left plot}) we show the temporal
evolution of $\left\langle W_{r}\right\rangle (t)$, i.e., the cumulative
payoff up to time $t$, according to Equation \ref{Eq:Acumulado_p_fixo} for
five $p$-values. We can observe that case $p=1/2$ corresponds to the regime
which $p $ changes with offer (\ref{Eq. Ganho_dep_oferta}).

In the same Figure (\textbf{right plot}) we show the behavior of the
variance of the payoff for the same $p$-values (Eq. \ref{local_gain_variance}%
). The variance can increase or decrease, respectively, for $p>1/2$ and $%
p<1/2$. For $p=1/2$ we have also the agreement with the case with $p$
dependent on the offer given by Equation \ref{Eq.local_dependent_offer_var}

\begin{figure}[th]
\begin{center}
\includegraphics[width=\columnwidth]{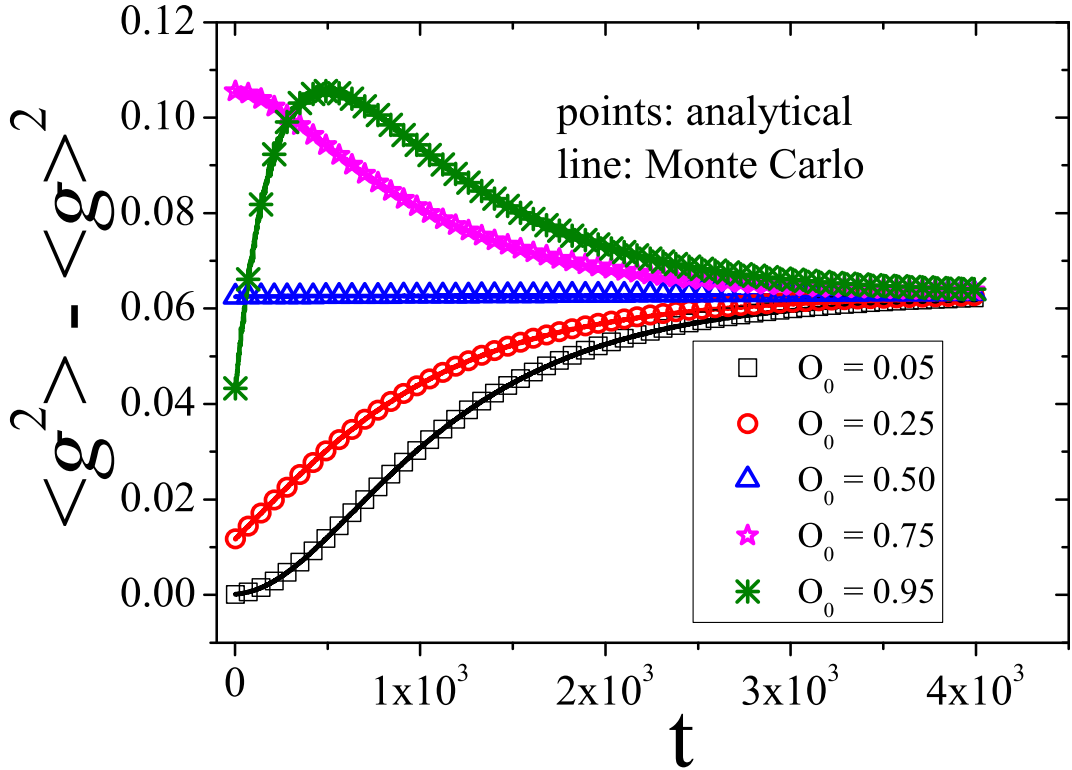}
\end{center}
\caption{Temporal evolution of the payoff dispersion for the case which
accepting depends on the offer. The points correspond to analytical results
and the lines to Monte Carlo simulations. }
\label{Fig:variancia_local_dep_offer}
\end{figure}

Now, it is interesting to observe what happens with other initial conditions
for variance of payoff when the acceptance depends on the offer. We can
observe in Fig. \ref{Fig:variancia_local_dep_offer} that values of payoff
dispersion always converge to the same value $var(g)\approx 0.06$ which does
not depend on $O_{0}$.

Particularly for $O_{0}=0.95$, we observe that variance has a maximum before
it converges to the steady state. This rich behavior is obviously related to
the fact that on average that offer converges $\left\langle
O_{t}\right\rangle \rightarrow O_{\infty }=1/2$ (see Eq. \ref%
{Eq:oferta_assintotico}). In this same plot, we also show that MC
simulations corroborate our analytical results.

So our reactive ultimatum game in mean field regime (two individuals
iteratively playing) and with the values averaged for a huge number of
repetitions (mean-field regime) is able to reproduce the intuitive aspect of
the ultimatum game, which corroborates real situations.\footnote{%
As seen in this simple video: https://www.youtube.com/watch?v=YXfEv-xEWtE}
Finally looking at the variance of the cumulative payoff we can also
estimate the value $\Phi (t)$ for this case as we performed for the $p$%
-fixed approach.

\begin{figure}[th]
\begin{center}
\includegraphics[width=0.5\columnwidth]{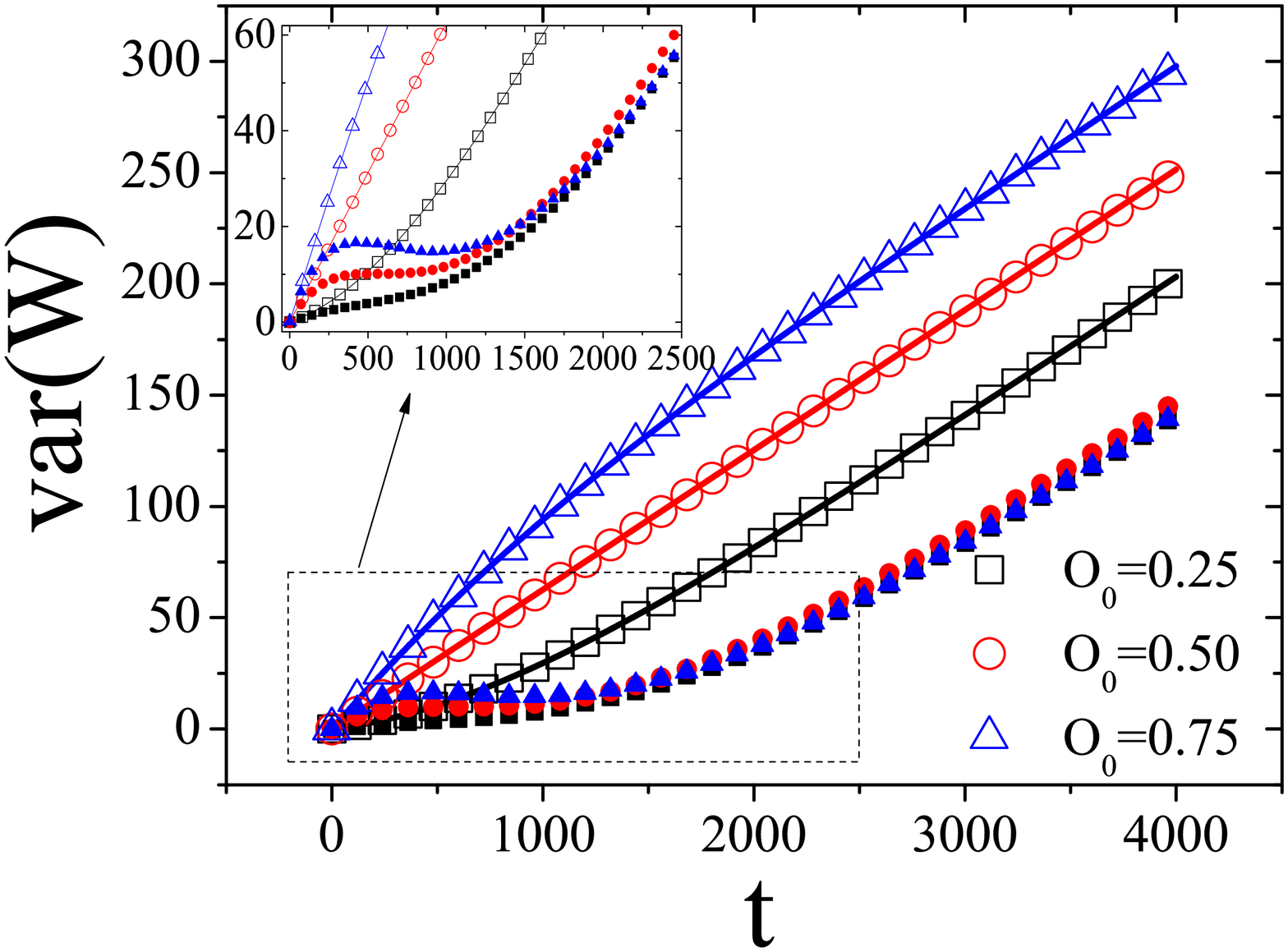}%
\includegraphics[width=0.5\columnwidth]{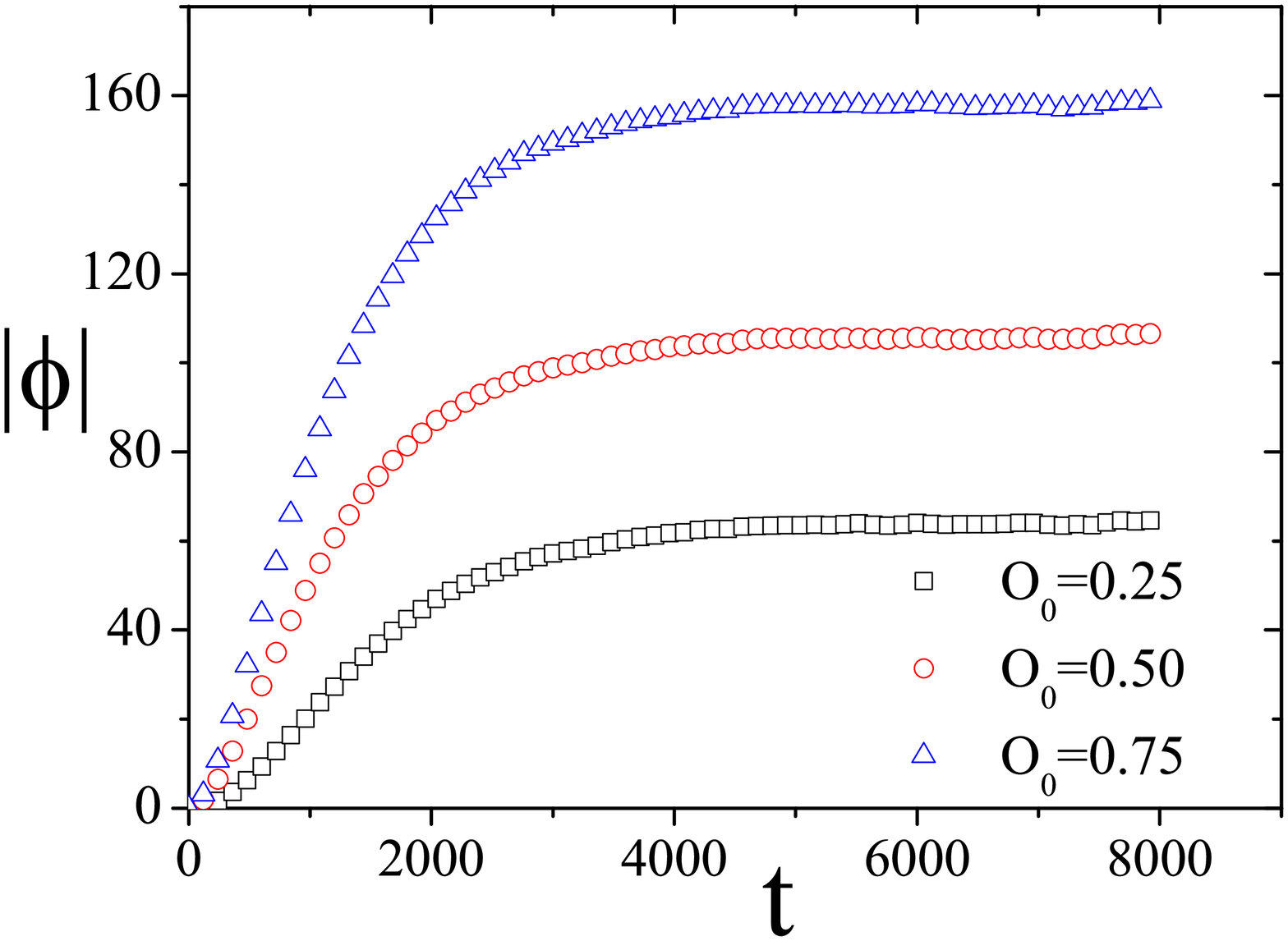}
\end{center}
\caption{{}\textbf{Left}: Variance of the cumulative payoff (filled in
points) and the sum of payoff variances (empty points) obtained by MC
simulations. The lines correspond to analytical results obtained by
Mean-Field approximation. \textbf{Right}: Temporal evolution of $\Phi (t)$
corresponding to results of the left plot in this same figure.}
\label{Fig:correlacao_dependente_oferta}
\end{figure}

In Fig. \ref{Fig:correlacao_dependente_oferta} (left) we show the variance
of the cumulative payoff $var(W_{r})$ as a function of time which is only
obtained by Monte Carlo simulations (full points). On the other hand, $%
\overline{var(W_{r})}$ (lines), the sum of payoff variances for all times $%
t^{\prime }<t$, were analytically estimated by Equations \ref{Eq.
Desc_dep_offer}, \ref{Eq:O2}, and \ref{Eq:offer_depends_on_time}.

The empty points correspond to $\overline{var(W_{r})}$ obtained by Monte
Carlo simulations. By performing the difference $var(W_{r})-$ $\overline{%
var(W_{r})}$ we obtain $\Phi (t)$. In Fig. \ref%
{Fig:correlacao_dependente_oferta} (right) we can observe that $\Phi (t)$
converges to a steady state well defined as the payoff and its variance.
Remember that this is different for the $p$-fixed approach (as seen in Fig. %
\ref{Fig:fi_with_fixed_p}).

In summary, we observed that offer-dependent acceptance produces a fair
steady state for these offers contrary to the expected rational behavior.
But in this version of reactive ultimatum game other important questions can
be answered: which are the effects of topologies, randomness, and the
neighborhood size on the offers. In the next section (second part of our
results) we analyze such effects on the reactive ultimatum game when the
acceptance depends on the offer.

\section{Results Part II: Coordination ($k\neq 1$) and Randomness Effects ($%
p\neq 0$)}

\label{Section:Results-II}

In this section we analyze the reactive ultimatum game in networks. We
initially concentrate our attention for populations that play under policy
I, in a regular graph with $k=4$. In this case, we can set $k=4 $ in
equation \ref{Eq:Exact_recurrence}, we have $\left\langle
O_{t+1}\right\rangle =\left\langle O_{t}\right\rangle +\ (\ \allowbreak
6\left\langle O_{t}\right\rangle ^{4}-8\left\langle O_{t}\right\rangle
^{3}+1)\varepsilon $. For $\left\langle O_{0}\right\rangle =O_{0}$. We can
iterate this equation. Simultaneously we have performed Monte Carlo
simulations in a square lattice by considering that a player will make an
offer to their four different neighbors and therefore will be the responder
to another four different neighbors. The player changes her decision with
respect to the offer only after having played with all neighbors, and the
synchronous our asynchronous version of the MC simulations which are similar
in this case.

\begin{figure}[th]
\begin{center}
\includegraphics[width=0.5%
\columnwidth]{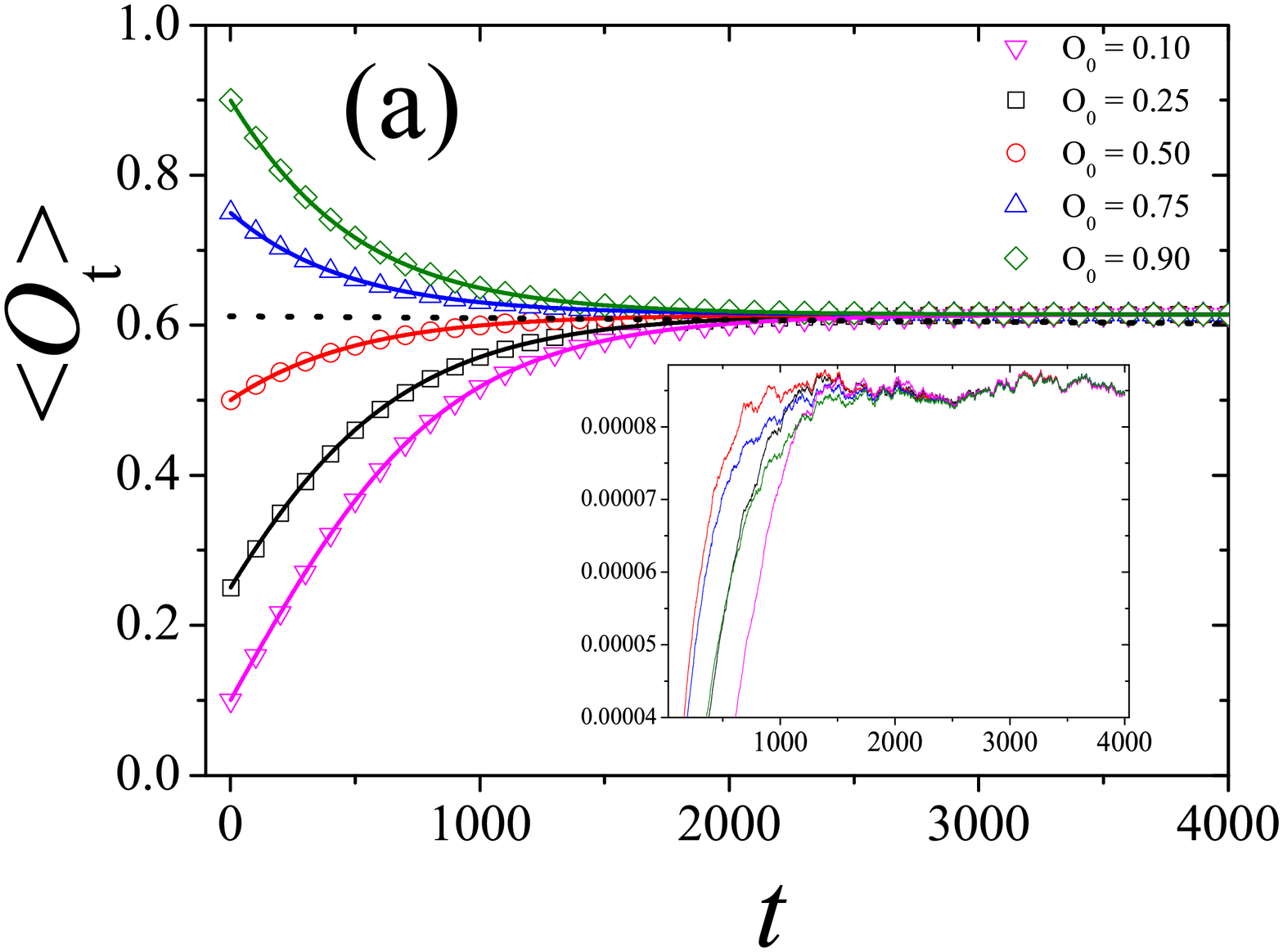}%
\includegraphics[width=0.5\columnwidth]{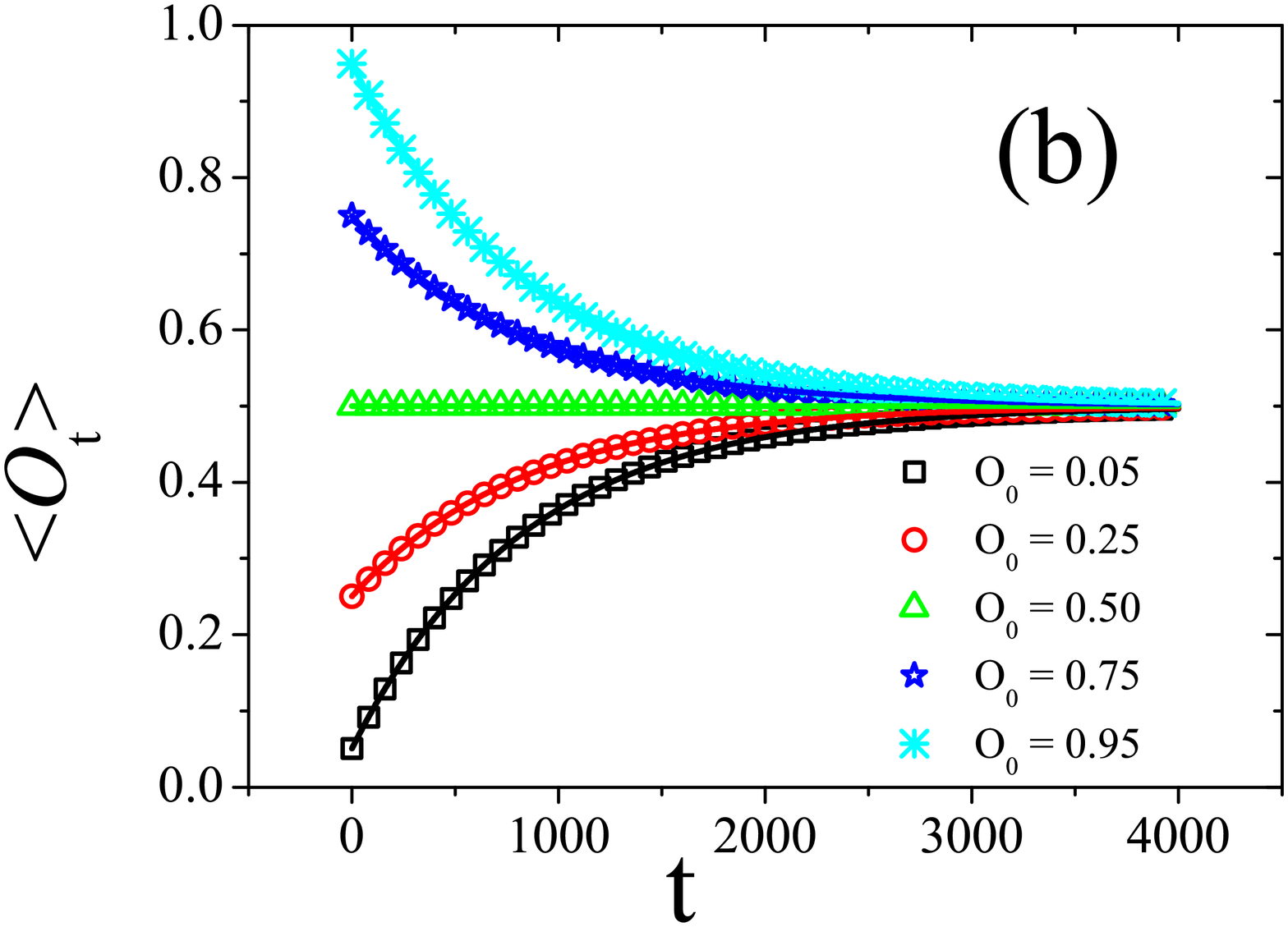} %
\includegraphics[width=0.5\columnwidth]{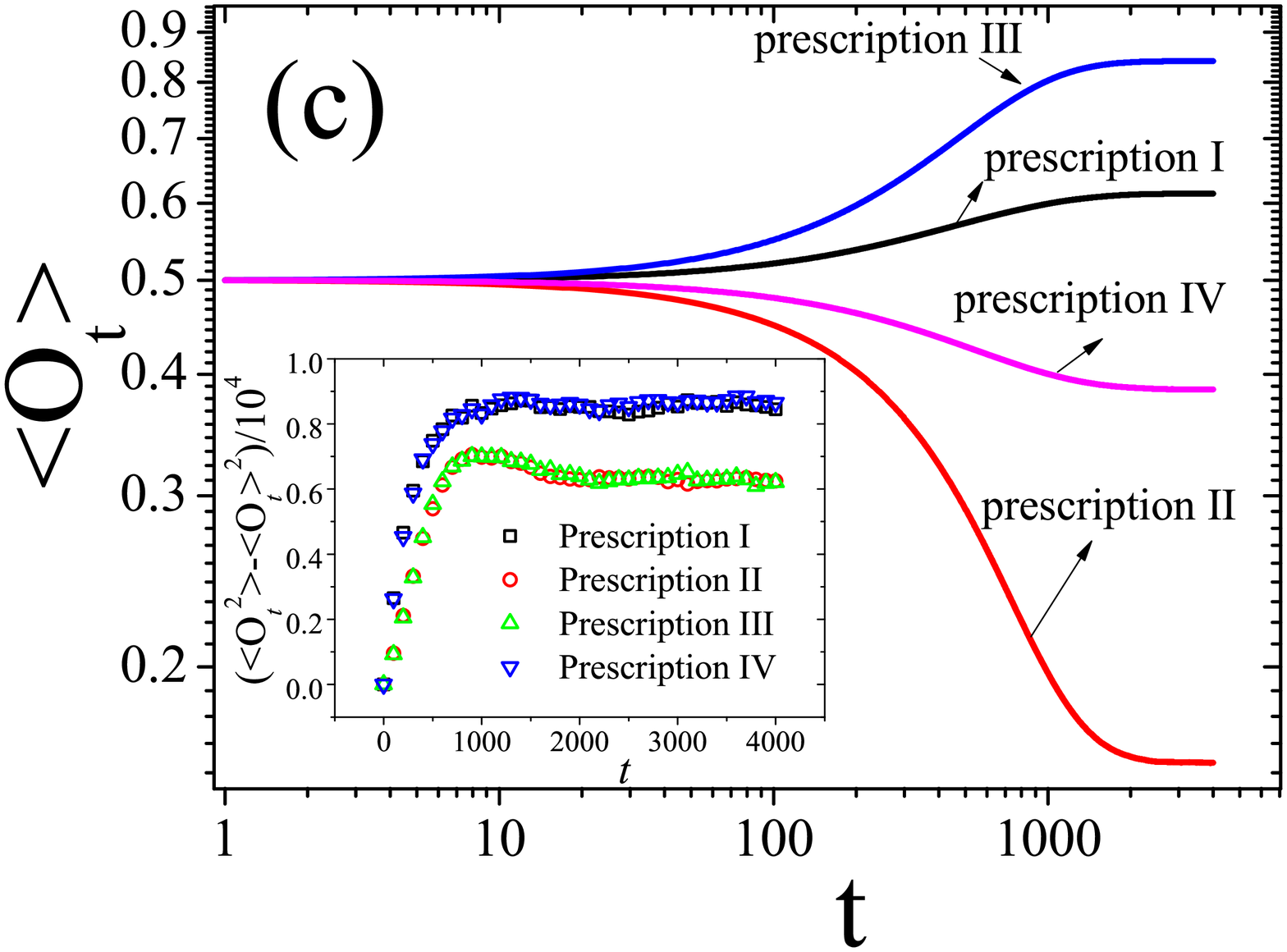}%
\includegraphics[width=0.5%
\columnwidth]{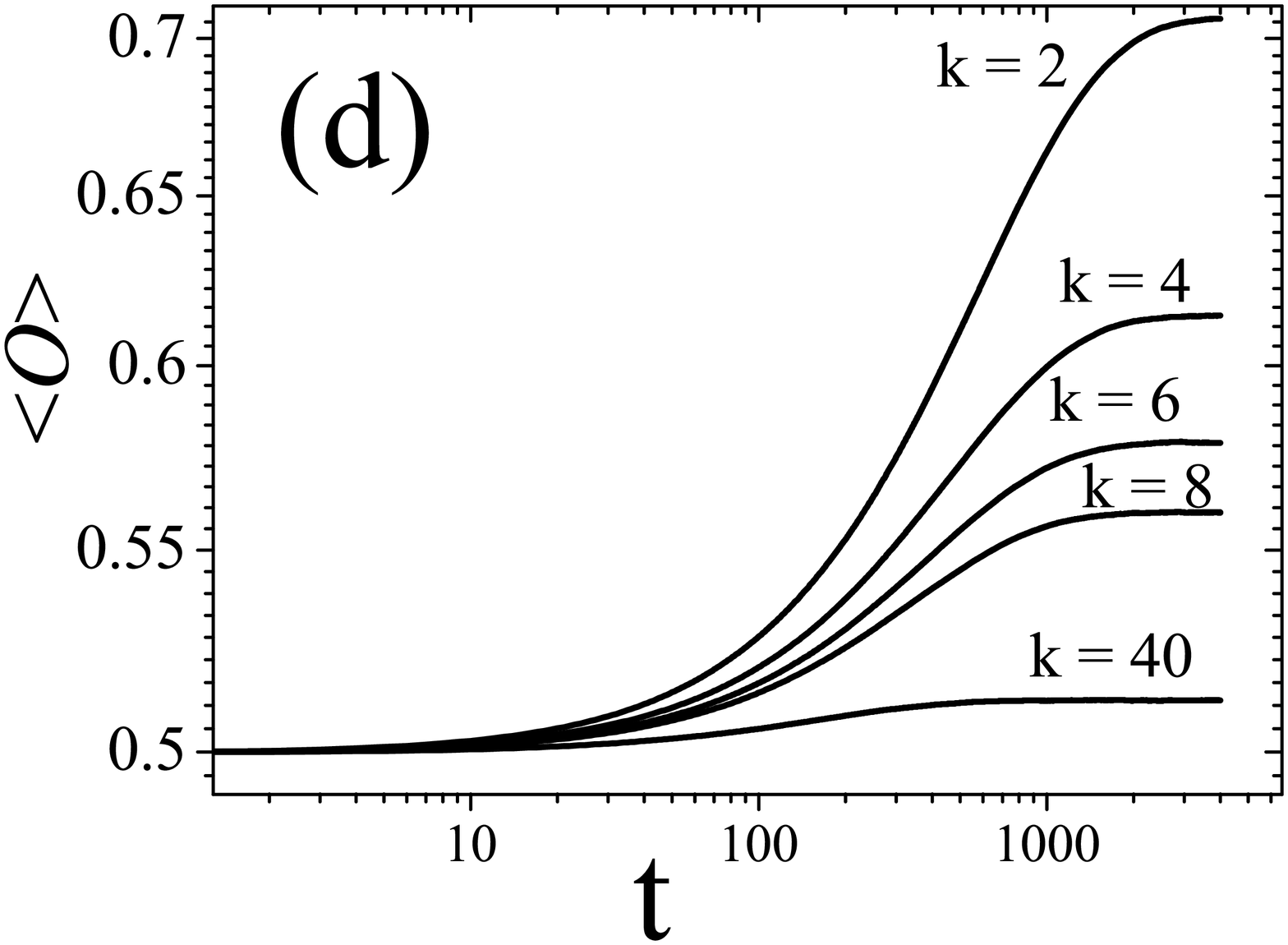}
\end{center}
\caption{(a): Average offer, for different initial value $O_{0}$, equally
attributed to all players. We considered the exact recurrence from Eq: 
\protect\ref{Eq:Exact_recurrence} (continuous curves) and MC simulation
(points) in a square lattice which corresponds to use $k=4$ in formula.
Inset plot correspond to stationarity of the offer dispersion. (b): Results
obtained from mean field which corresponds to $k=1$ (no policy dependent).
(c): Results for the other policies also for $k=4$. (d) Results for policy I
simulated on the square but with arbitrary neighborhood for different $k$%
-values. }
\label{Fig:stationary_state}
\end{figure}

Here an important question to ask is about the influence of randomness on
these results. If we imagine for example a small world built from a simple
ring or even a square lattice with coordination $k_{0}$, by introducing a
rewiring probability $p$, we have $\left\langle k\right\rangle =k_{0}$ but
the result corresponds exactly, for example to policy I, by changing $k$ by
corresponding $\left\langle k\right\rangle $? This is not what happens. This
occurs only if $k_{0}$ is large; for smaller $k_{0}$ we have a dependence on 
$p$ as can be observed in the color maps of Figure \ref%
{Fig:Randomness_effects}.

\begin{figure}[th]
\begin{center}
\includegraphics[width=0.5\columnwidth]{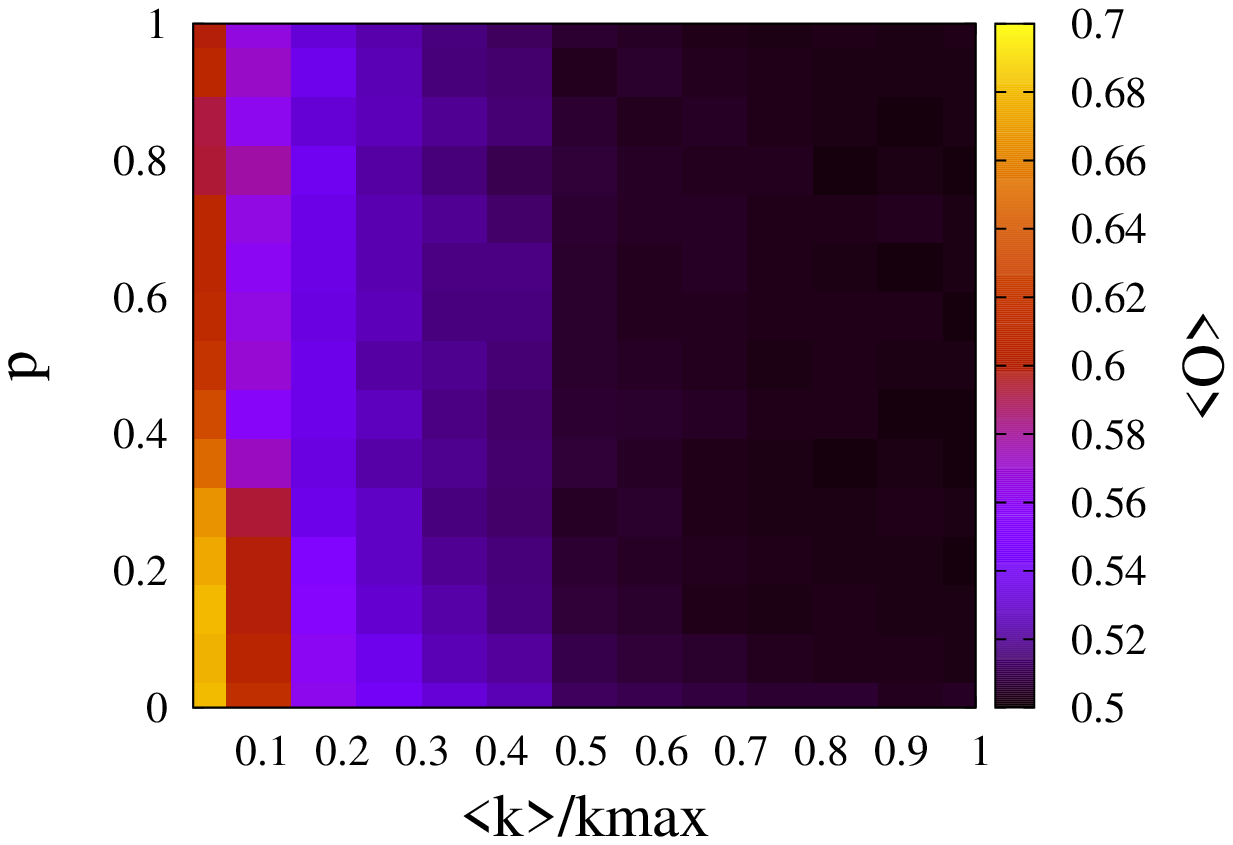}\includegraphics[width=0.5%
\columnwidth]{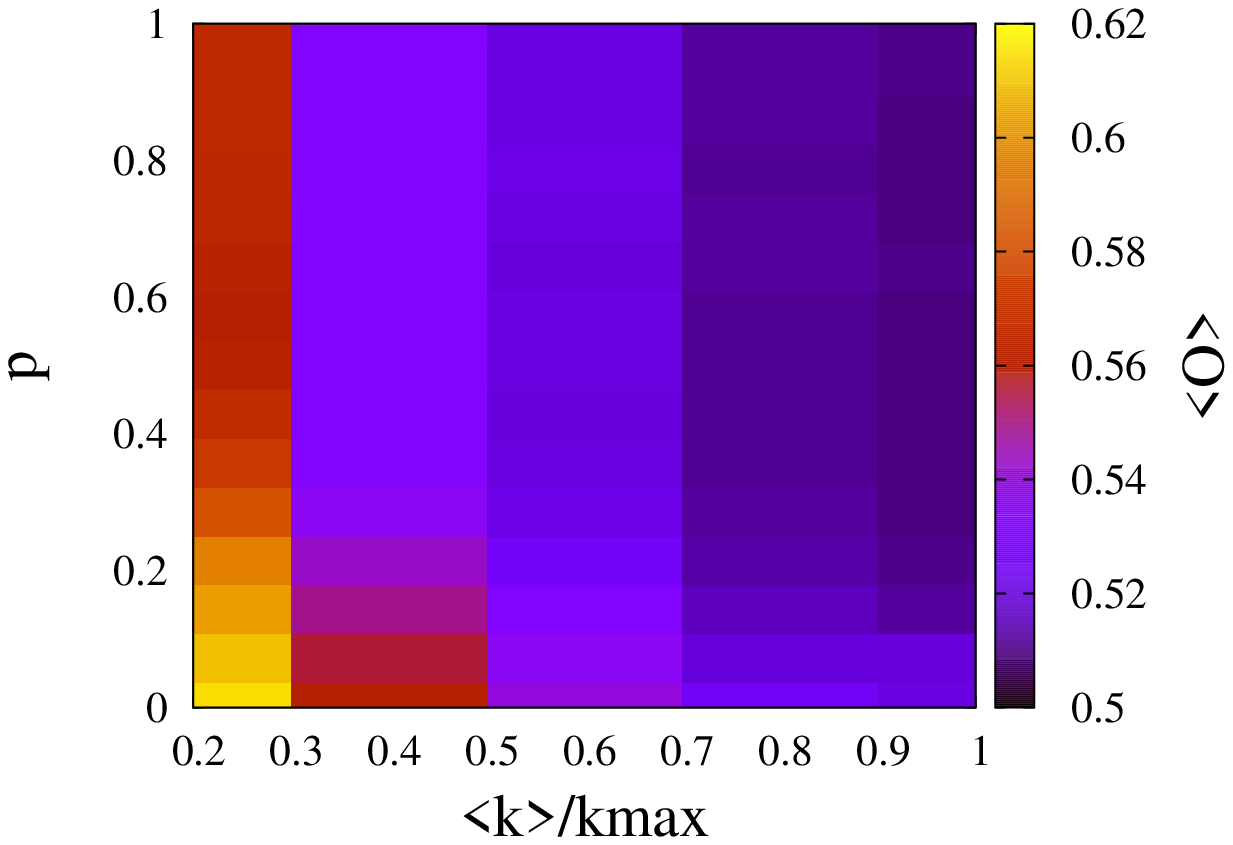}
\end{center}
\caption{Randomness effects on stationary offers for the policy I: a) Left
plot: small world built from a ring b) Right plot: small world built from a
square lattice. }
\label{Fig:Randomness_effects}
\end{figure}
Such behavior can be checked by looking the dependence of stationary offer
as function of $p$ and $\left\langle k\right\rangle $ as shown in fig. \ref%
{Fig:Randomness_effects}. We performed simulations in a small world starting
from a ring and a square lattice. It is interesting observe that for low
coordination even for $p=1$ we do not obtain the result expected for mean
field ($\left\langle O_{t\rightarrow \infty }\right\rangle =1/2$).

Now it is interesting to analyze the effects about the payoff of the players
in populations under four different policies. We want to show the effects
about distribution of payoff in populations as a function of time
considering the populations in which the offers are performed under 4
different policies and acceptance occurs with probability exactly the offer
of the player. We consider ($k=4$). We can see from plot (C) in Fig. \ref%
{Fig:stationary_state} that policy 3 leads to higher offers. This happens
because players with this behavior only decrease their offers in really
favorable situation; they prefer to deal with more players under lower offer
values than playing with only one player under higher offer values. Bigger
offers mean higher acceptance probabilities, which mean larger number of
deals.

By considering the payoff obtained by players in populations interacting
under the different policies, we analyzed statistics related to payoff
obtained by the players in populations interacting under these different
policies separately. We consider $k=4$, for the sake of simplicity. One
question to ask is how the payoff is distributed among the players along
time. In this case, we can use a interesting concept from Economics, the
Gini coefficient. Considering that $N$ players have their payoffs at time $t$
in increasing order: $g_{1}(t)\leq g_{2}(t)\leq ....\leq g_{N}(t)$.$\ $So we
consider the cumulative distribution: 
\begin{equation*}
\varphi _{i}(t)=\frac{\sum_{j=1}^{i}g_{j}(t)}{\sum_{j=1}^{N}g_{j}(t)}
\end{equation*}

The Lorentz ($\varphi _{i}(t)\times i/N$) curve shows the corresponding
wealth (sum of payoffs) corresponding to population fraction $f_{i}=i/N$. We
expect an identity function for a well distributed payoff. By a trapezoidal
formula, the Gini coefficient can be estimated by

\begin{equation*}
G(t)=1-\frac{1}{N}\sum_{i=1}^{N}\varphi _{i}(t)
\end{equation*}%
which measures the difference between the Lorentz curve and the identity
function. This number changes from 0 up to 1, and the higher the value of $G$%
, the worse is the payoff distribution.

\begin{figure}[th]
\begin{center}
\includegraphics[width=0.5\columnwidth]{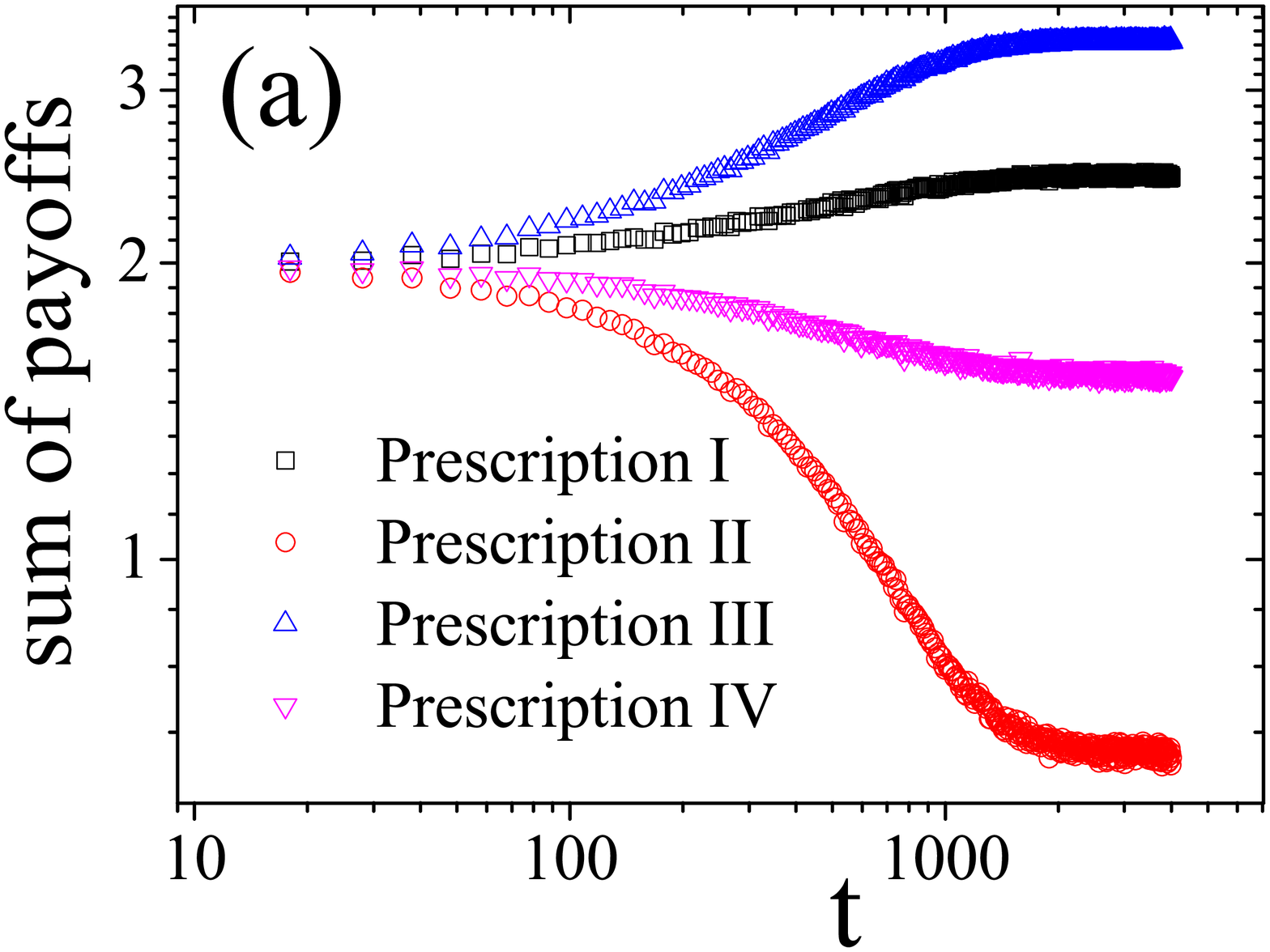}%
\includegraphics[width=0.5\columnwidth]{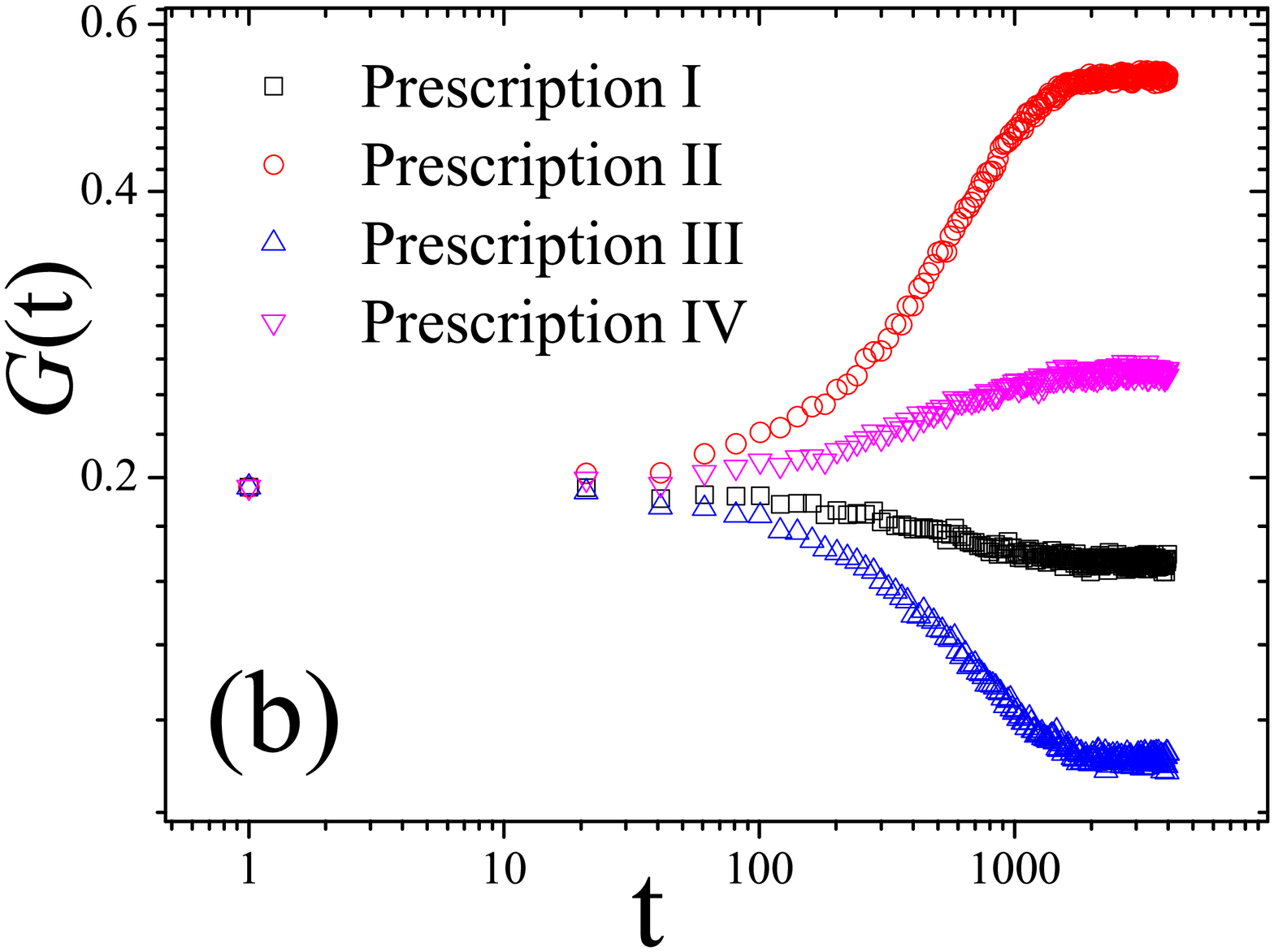}
\end{center}
\caption{(a): Average total payoff of the players for $k=4$ by considering
the different policies. (b): Corresponding Gini coefficient of the average
payoff described by Plot (a). }
\label{Fig:payoff_and_Gini}
\end{figure}

Since we analyzed the properties of populations under different policies for 
$k=4$, now we would like to better explore a general formula for the
stationary offer for arbitrary coordination, considering populations under
proportions of different policies. If we consider that $\rho _{c}$, $\rho
_{G}$, $\rho _{HC}$ and $\rho _{M}$ are the densities of conservative,
greedy, highly conservative and moderated players, we can write that with
the players inserted in a population with coordination $k$, is%
\begin{equation*}
\begin{array}{ccc}
\left\langle O_{t+1}\right\rangle & \approx & \left\langle
O_{t}\right\rangle +\rho _{c}\cdot \left[ 2\Pr (0\leq n_{a}\leq
k/2|\left\langle O_{t}\right\rangle )-1\right] \varepsilon +\rho _{G}\cdot %
\left[ 2\Pr (n_{a}=0|\left\langle O_{t}\right\rangle )-1\right] \varepsilon +
\\ 
&  &  \\ 
&  & +\ \rho _{HC}\left[ 2\Pr (0<n_{a}<k|\left\langle O_{t}\right\rangle )-1%
\right] \varepsilon +\rho _{M}\left[ 2\Pr (0<n_{a}<k/2-1|\left\langle
O_{t}\right\rangle )-1\right] \varepsilon%
\end{array}%
\end{equation*}
which results in

\begin{equation}
\begin{array}{lll}
\left\langle O_{t+1}\right\rangle & \approx & \left\langle
O_{t}\right\rangle +\rho _{c}\left( 2\sum_{m=0}^{k/2}\frac{k!\left\langle
O_{t}\right\rangle ^{m}(1-\left\langle O_{t}\right\rangle )^{k-m}}{m!(k-m)!}%
-1\right) +\rho _{G}\left( 2(1-\left\langle O_{t}\right\rangle
)^{k}-1\right) +\rho _{HC}\left( 1-2\left\langle O_{t}\right\rangle
^{k}\right) \\ 
&  &  \\ 
&  & +\rho _{M}\left( 2\sum_{m=0}^{k/2-1}\frac{k!\left\langle
O_{t}\right\rangle ^{m}(1-\left\langle O_{t}\right\rangle )^{k-m}}{m!(k-m)!}%
-1\right) \varepsilon%
\end{array}
\label{Eq: general}
\end{equation}

Obviously, Eq. \ref{Eq:Exact_recurrence} is a particular case of Eq. \ref%
{Eq: general} ($\rho _{c}=1$, $\rho _{G}=\rho _{HC}=\rho _{M}=0$). So our
work now is to change the proportions $\rho _{c}$, $\rho _{G}$, $\rho _{HC}$
and $\rho _{M}$, by numerically solving this equation and answering an
important question: Is there some proportion that is able to change the
behavior as $\left\langle O_{t\rightarrow \infty }\right\rangle =1/2$? First
of all, it is important to mention that all results obtained by numerical
integration of Eq. \ref{Eq: general} were checked by performing simulations
in rings and square lattices with arbitrary coordination. For this reason we
will omit any information about MC simulations from that part until the
final results, but remember that we have a perfect agreement between MC
simulations and numerical integration of Eq. \ref{Eq: general}.

\begin{figure}[th]
\begin{center}
\includegraphics[width=0.5\columnwidth]{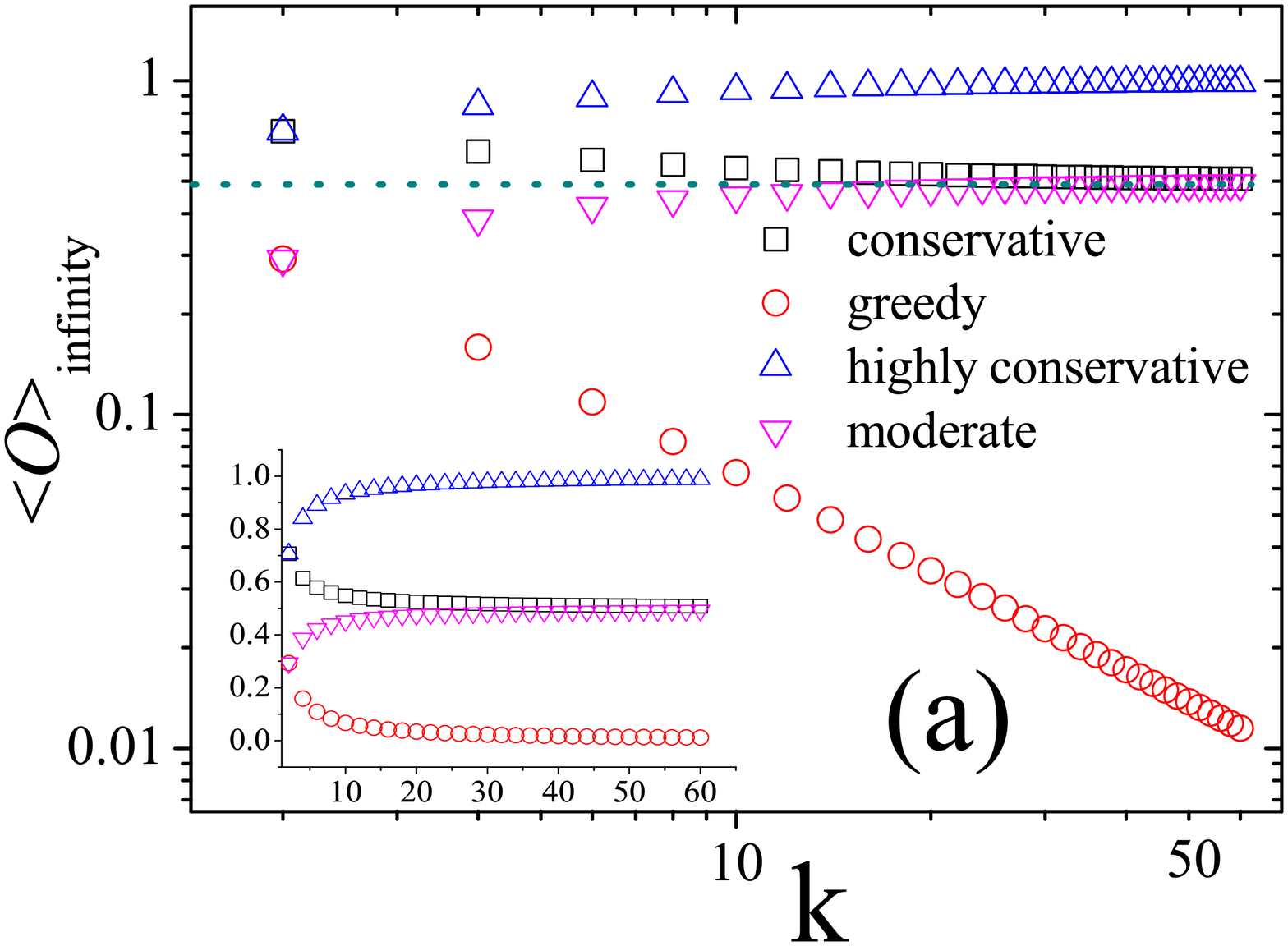}%
\includegraphics[width=0.5\columnwidth]{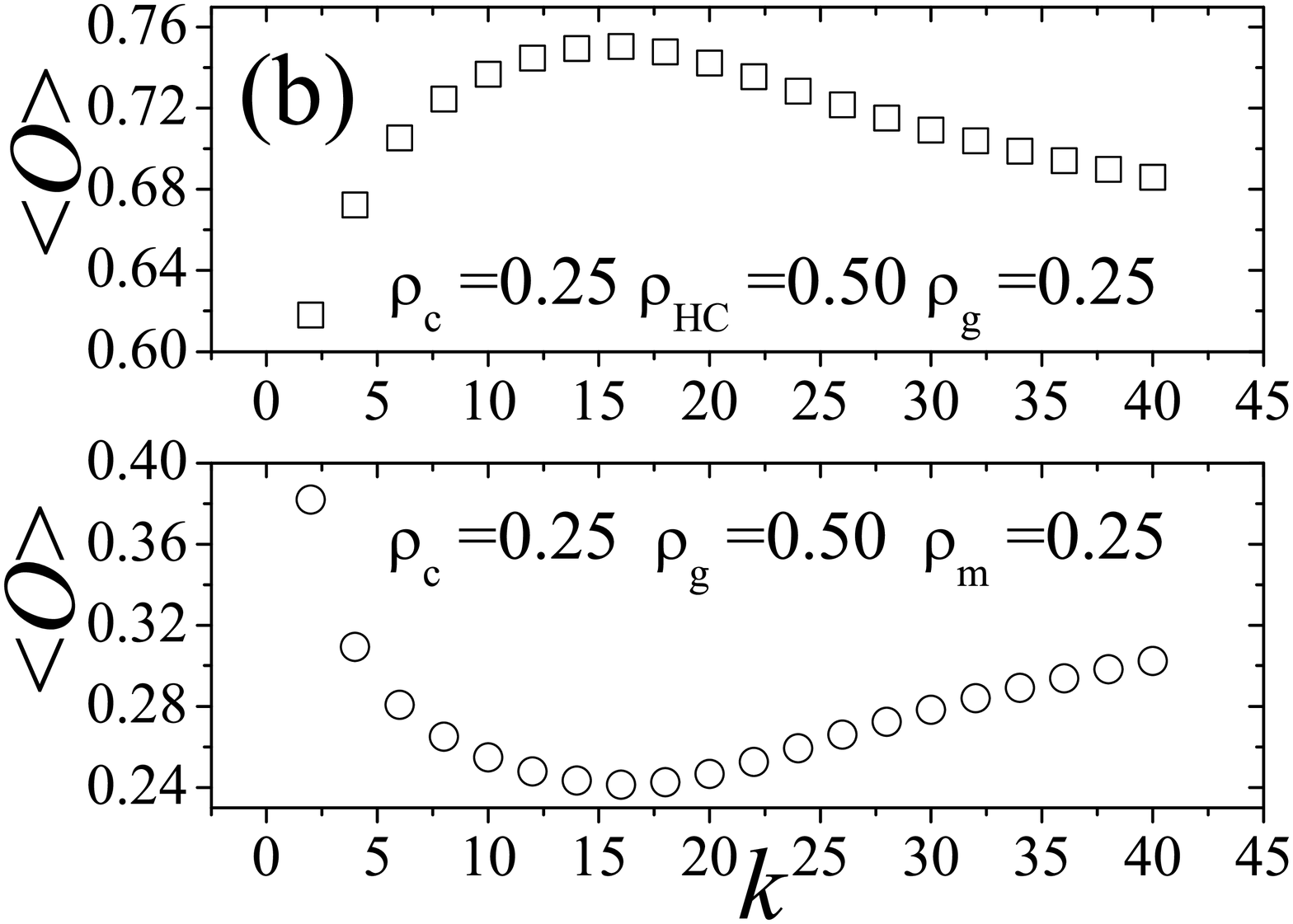}
\end{center}
\caption{\textbf{(a)}: Stationary average offers as a function of $k$
considering homogeneous populations - players follow a single policy - in
log-log scale. \textbf{(b)}: Some mixing of policies (heterogeneous
population) illustrating one case where there is a coordination that
maximizes the stationary offer and one that minimizes it. }
\label{Fig:heterogenous}
\end{figure}

First, we would like to analyze the stationary average offer for mixing of
different strategies, by looking at differences between the homogeneous
populations (i.e., that one where players only use the same policy (I, II,
III, or IV). In plot (a), Fig. \ref{Fig:heterogenous}, we show the behavior
of $\left\langle O_{t\rightarrow \infty }\right\rangle $ as a function of $k$
in log-log scale. Each plot corresponds to one population interacting
according to a specific policy (we denote it as homogeneous population). The
inset plot corresponds to the same plot in linear scale. We can observe that
in policies I and IV, the stationary offers converge to $1/2$ when $%
k\rightarrow \infty $, differently from II and III.

It is important to mention that a population with only greedy players leads
to an algebraic decay of the offer as coordination ($k$): $\left\langle
O_{t\rightarrow \infty }\right\rangle \sim k^{-\xi }$. We measured the
exponent $\xi \approx 0.95$ by using $k_{\max }=40$ and $\xi \approx 0.97$
for $k_{\max }=60$ which indicates a kind of hyperbolic scaling in
coordination $\left\langle O_{t\rightarrow \infty }\right\rangle \sim 1/k$.

The first experiment with heterogeneous population keeps ($\rho _{c}=\rho
_{G}=\rho _{HC}=\rho _{M}=1/4)$. In this case, surprisingly the stationary
case is $\left\langle O_{t\rightarrow \infty }\right\rangle =1/2$
independently from $k$ (by simplicity we omit this obvious plot). We cannot
observe such behavior in the studied homogeneous populations.

Other exotic choices can be performed in which $\left\langle O_{t\rightarrow
\infty }\right\rangle $ shows convex and concave behaviors as function of $k$%
, i.e., with extrema well defined as we can observe in plot (b), Fig. \ref%
{Fig:heterogenous}.

\section{Conclusions}

In this paper we analyze some important aspects of populations which
interact under a reactive ultimatum game. First we extended results of a
recent publication where acceptance of the offers occurs with fixed
probability $p$. We show an interesting behavior for the sum of all temporal
correlations of the payoff from $t^{\prime }=0,...,t$: $\Phi (t)$, which
changes its signal in time that depends on the acceptance probability $p$,
that is a property from the fact that $\left\langle O_{t}\right\rangle $
increases (respectively, decreases) as $p>0$ (resp.$,<0$) as function of
time.

Based on the fact that unfair offers have small acceptance probabilities, we
proposed a new model where acceptance occurs with probability $O_{t}$, i.e.
the offer of the opponent. In this case a mean field regime leads to a
interesting stationary fair offer: $O_{t\rightarrow \infty }=1/2$
independently from the initial offer $O_{0}$. Thus, the sum of the temporal
correlations of the payoff has a steady state well defined, but depends on $%
O_{0}$.

When studied in networks the model does not present $O_{t\rightarrow \infty
}=1/2$ for low coordination (small $k$) whatever the policy analyzed.
Particularly for $k=4$ we showed that the average payoff is larger and the
Gini coefficient is smaller for the policy that decreases the respective
offer only when all players have accepted the offer at hand. This apparently
altruistic player gains low values as proposer, but higher values as a
responder; this combination leads to a well distributed payoff. We show that
the absolutely greedy policy (II) leads to low payoffs and to high Gini
coefficients.

Further, we introduced four policies that differ in how each player
increases/decreases her offer. Only two policies present $\left\langle
O_{t\rightarrow \infty }\right\rangle =1/2$ for $k\rightarrow \infty $.
However a perfect equilibrium among policies, i.e. 1/4 of population for
each policy, leads to $\left\langle O_{t\rightarrow \infty }\right\rangle
=1/2$ independently from $k$. There is a breaking of monotonicity of $%
\left\langle O_{t\rightarrow \infty }\right\rangle (k)$ for mixing of
strategies, which presents $k$-values where $\left\langle O_{t\rightarrow
\infty }\right\rangle $ is a extreme, either maximum or minimum value.

\bigskip

\bigskip


\begin{thebibliography}{99}
\bibitem{Neumann} J. von Neumann, O. Morgenstern, "Theory of Games and
Economic Behavior" (Princeton University Press, Princeton NJ) (1944)

\bibitem{Smith} J. M. Smith, Journ. Theor. Biol. \textbf{47},\ 209-221 (1974)

\bibitem{SzaboFath} G. Szab\'{o}, G. F\'{a}th, Phys. Rep. \textbf{446},
97-216(2007)

\bibitem{Guth1982} W. Guth, R. Schmittberger and B. Schwarze, J. Econ.
Behav. Org., \textbf{24,} 153 (1982).

\bibitem{Henrich2000} J. Henrich, Am. Econ. Rev. \textbf{90}, 973 (2000)

\bibitem{juespacial} A. Szolnoki, M. Perc, G. Szab\'{o}, Phys. Rev. Lett. 
\textbf{109}, 078701-1 (2012).

\bibitem{estudoneural} A.G. Sanfey, J. K. Rilling, J.A. Aronson, L. E.
Nystrom and J. D. Cohen, Science \textbf{300} 1755 (2003).

\bibitem{Burnaham} T. C. Burnham, Proc. Royal Soc. B, \textbf{274},
2327-2330 (2007)

\bibitem{Nowak2000} M. A. Nowak, K. M. Page, K. Sigmund, Science, \textbf{289%
} 1773-1775 (2000).

\bibitem{Enock2014} E. Almeida, R. da Silva, A. S. Martinez, Physica A, 
\textbf{412}, 54 (2014)

\bibitem{SilvaJTB} R. da Silva, G. A. Kellermann and L. C. Lamb, J. Theor.
Biol., \textbf{258}, 208-218 (2009).

\bibitem{SIlvaBJP} R. da Silva and G. A. Kellerman, Braz. J. Phys., \textbf{%
37}, 1206-1211 (2007).
\end{thebibliography}
\end{document}